\documentclass[iop]{emulateapj}
\usepackage{hyperref}

\newcommand{\tmb}{$T_{\rm mb}~$}
\newcommand{\tmbns}{$T_{\rm mb}$}
\newcommand{\snu}{$S_{\nu}~$}

\newcommand{\tex}{$T_{\rm ex}~$}
\newcommand{\texns}{$T_{\rm ex}$}
\newcommand{\tk}{$T_{k}~$}

\newcommand{\dfrac}{$D_{\rm frac}^{\rm N_2H^+}~$}
\newcommand{\dfracns}{$D_{\rm frac}^{\rm N_2H^+}$}
\newcommand{\ntdp}{$\rm N_2D^+~$}
\newcommand{\ntdpns}{$\rm N_2D^+$}
\newcommand{\nthp}{$\rm N_2H^+~$}
\newcommand{\nthpns}{$\rm N_2H^+$}
\newcommand{\kms}{$\rm km~s^{-1}~$}
\newcommand{\kmsns}{$\rm km~s^{-1}$}
\newcommand{\nht}{$n_{\rm H_2}~$}

\newcommand{\opr}{OPR$^{\rm H_2}~$}

\newcommand{\oprz}{OPR$^{\rm H_2}_0~$}
\newcommand{\oprzns}{OPR$^{\rm H_2}_0$}
\newcommand{\ohtdp}{$\rm o$-$\rm H_2D^+~$}
\newcommand{\ohtdpns}{$\rm o$-$\rm H_2D^+$}
\newcommand{\ffd}{$f_{D}~$}

\newcommand{\ffdz}{$f_{D,0}~$}
\newcommand{\ffdzns}{$f_{D,0}$}
\newcommand{\aff}{$\alpha_{\rm ff}~$}
\newcommand{\affns}{$\alpha_{\rm ff}$}
\newcommand{\nh}{$n_{\rm H}~$}
\newcommand{\nhns}{$n_{\rm H}$}
\newcommand{\nho}{$n_{\rm H,1}~$}
\newcommand{\nhons}{$n_{\rm H,1}$}
\newcommand{\nhz}{$n_{\rm H,0}~$}
\newcommand{\nhzns}{$n_{\rm H,0}$}

\shorttitle{The Deuterium Fraction in Massive Starless Cores and Dynamical Implications}
\shortauthors{Kong et al.}

\begin{document}

\title{The Deuterium Fraction in Massive Starless Cores and Dynamical Implications}

\author{Shuo Kong\altaffilmark{1}}
\affil{Dept. of Astronomy, University of Florida, Gainesville, Florida 32611, USA}

\author{Jonathan C. Tan\altaffilmark{1,2}}
\affil{Dept. of Astronomy, University of Florida, Gainesville, Florida 32611, USA}
\affil{Dept. of Physics, University of Florida, Gainesville, Florida 32611, USA}

\author{Paola Caselli\altaffilmark{3}}
\affil{Max-Planck-Institute for Extraterrestrial Physics (MPE), Giessenbachstr. 1, D-85748 Garching, Germany}

\author{Francesco Fontani\altaffilmark{4}}
\affil{INAF - Osservatorio AstroÞsico di Arcetri, I-50125, Florence, Italy}

\author{Thushara Pillai\altaffilmark{5,6}}
\affil{California Institute of Technology, Cahill Center for Astronomy and Astrophysics, Pasadena, CA 91125, USA}
\affil{Max Planck Institut f\"{u}r Radioastronomie, Auf dem H\"{u}gel 69, D-53121 Bonn, Germany}
%jct*** - add Pillai here

\author{Michael J. Butler\altaffilmark{7}}
\affil{Max Planck Institute for Astronomy, K\"onigstuhl 17, 69117 Heidelberg, Germany}

\author{Yoshito Shimajiri\altaffilmark{8}}
\affil{Laboratoire AIM, CEA/DSM-CNRS-Universit\`e Paris Diderot, IRFU/Service d' Astrophysique, CEA Saclay, 91191 Gif-sur-Y vette, France}

\author{Fumitaka Nakamura\altaffilmark{9}}
\affil{National Astronomical Observatory of Japan, 2-21-1 Osawa, Mitaka, 181-8588 Tokyo, Japan}

\author{Takeshi Sakai\altaffilmark{10}}
\affil{Graduate School of Informatics and Engineering, The University of Electro-Communications, Chofu, Tokyo 182-8585, Japan}

\begin{abstract}
We study deuterium fractionation in two massive starless/early-stage cores C1-N
and C1-S in Infrared Dark Cloud (IRDC) G028.37+00.07, first identified
by \citet{2013ApJ...779...96T} with {\it ALMA}.  Line emission from
multiple transitions of $\rm N_2H^+$ and $\rm N_2D^+$ were observed
with the {\it ALMA}, {\it CARMA}, {\it SMA}, {\it JCMT}, {\it NRO 45m}
and {\it IRAM 30m} telescopes. By simultaneously fitting the spectra,
we estimate the excitation conditions and deuterium fraction,
\dfracns$\equiv [\rm N_2D^+]/[N_2H^+]$, with values of
\dfracns$\simeq0.2$--$0.7$, several orders of magnitude above the
cosmic [D]/[H] ratio. Additional observations of o-H$_2$D$^+$ are also
presented that help constrain the ortho-to-para ratio of $\rm H_2$,
which is a key quantity affecting the degree of deuteration. We then
present chemodynamical modeling of the two cores,
exploring especially the implications for the collapse rate relative
to free-fall, $\alpha_{\rm ff}$.
In order to reach the high level of observed deuteration of \nthpns,
we find that the most likely evolutionary history of the cores
involves collapse at a relatively slow rate, $\lesssim1/10$th of
free-fall.
\end{abstract}


\section{Introduction}

Massive stars produce powerful feedback that
helps to shape the structure of galaxies and even the intergalactic
medium. However, the formation of massive stars still involves many
open questions, in part because 
the initial conditions of massive star birth are relatively rare,
distant and deeply embedded in massive clump/protocluster
envelopes. Infrared Dark Clouds (IRDCs) are promising places to search
for these initial conditions since they contain large quantities of
cold ($\sim$10~K), high density ($n_{\rm H}\gtrsim 10^5\:$cm$^{-3}$)
gas (e.g., \citealt{2006ApJ...641..389R}; \citealt{2006A&A...450..569P};
\citealt{2009ApJ...696..484B,2012ApJ...754....5B}; 
see review by \citealt{2014prpl.conf..149T}, hereafter T14).

Theoretically, one of the key questions is whether the formation
mechanism of massive stars is a scaled-up version of low-mass star
formation \citep{1987ARA&A..25...23S}
or not.  Two main competing models of massive star formation have been
put forward, one is ``Turbulent Core Accretion" \citep{2002Natur.416...59M,2003ApJ...585..850M},
which is a scaled-up version of core accretion models for low-mass
star formation. The other is ``Competitive Accretion" (\citealt{2001MNRAS.323..785B};
see also \citealt{2010ApJ...709...27W}). These two models involve very different
initial conditions and accretion mechanisms. Turbulent Core Accretion
assumes a near-virialized massive starless core for the initial
condition, while Competitive Accretion forms a massive star at the
center of a globally collapsing clump that fragments into a swarm of
low-mass protostars. To test between the two models, it is critical to
identify and characterize massive starless cores.

Once identified, it is then important to measure the virial state of a
core to understand its dynamical state. The \citet{2001MNRAS.323..785B}
model of Competitive Accretion involves a gas cloud that is undergoing
rapid collapse from a ``sub-virial'' state.
One obstacle to determining the virial state is estimating the
strength of magnetic fields. Strong magnetic fields could provide
significant support in addition to other sources (mainly turbulence,
since thermal pressure is dynamically unimportant in the massive, cold
structures of IRDCs). However, while there is evidence for strong
$B$-fields around massive protostars \citep[e.g.,][]{2009Sci...324.1408G,2014ApJ...792..116Z}, 
there are very few measurements at earlier
stages. Recently, \citet{2015ApJ...799...74P} have presented the first
measurement of $B$-field strengths in dark, presumably starless
regions of IRDCs, finding evidence for dynamically strong field
strengths.

As an alternative approach, in this paper we try and assess the age of
a core by astrochemical indicators, in particular the level of
deuteration of key species.
We compare the chemical age of a core with its dynamical (sound
crossing or free-fall) timescale. If the chemical age is much greater
than the dynamical timescale, then we expect that the core must have
reached approximate virial equilibrium, so that if it is undergoing
collapse it is at a relatively slow rate, perhaps regulated by
magnetic field support.

The particular astrochemical indicator that we examine is the
deuterium fraction of $\rm N_2H^+$ ($D_{\rm frac}^{\rm N_2H^+} \equiv
{\rm [N_2D^+]/[N_2H^+]}$). It rises in the cold, dense conditions of
starless cores, increasing by $\sim$3-4 orders of magnitude.
Theoretically, this is due to the fact that the parent exothermic
reaction $\rm H_3^+ + HD \rightleftharpoons H_2D^+ + H_2 + 232~K$
\citep[all in the para state;][]{1992A&A...258..479P} is favored at
low temperatures ($\sim$10~K).  Observationally, \dfrac has been shown
to be a good evolutionary tracer for both low-mass and high-mass cores
\citep[see, e.g.,][]{2005ApJ...619..379C,Emprechtinger2009,
  2013ApJ...765...59F,2011A&A...529L...7F}. Indeed, it is probably the
best tracer of pre-stellar cores, e.g., compared to $D_{\rm
  frac}^{\rm HNC}$ and $D_{\rm frac}^{\rm NH_3}$
\citep{2015A&A...575A..87F}.  Thus, overall, we consider \ntdp to be
the best diagnostic tool for detecting massive starless cores given
the astrochemical model prediction of high abundance in cold, dense
regions.  Other methods, such as dust continuum
\citep[e.g.,][]{2006ApJ...641..389R}, dust extinction
\citep[e.g.,][]{2012ApJ...754....5B}, and other molecular line
observations (e.g., NH$_3$, \nthpns), are likely subject to
contamination from the much more massive clump envelope surrounding
the cores. Dust continuum emission is also in general more sensitive
to warmer, protostellar cores, rather than starless cores.

We have developed a chemical model \citep{2015ApJ...804...98K} to
describe the time evolution of \dfrac \citep[also
  see][]{2013A&A...551A..38P}, including for dynamical models of
collapsing cores.  Measurement of the abundances $\rm
[N_2D^+]~and~[N_2H^+]$, and thus \dfracns, in starless cores, allows
estimation of core age and so constrains the dynamical history of its
collapse, e.g., the collapse rate relative to free-fall.

Two massive starless/early-stage cores have been identified in IRDC
G028.37+00.07 \citep[kinematic distance of 5 kpc,][]{2006ApJ...639..227S},
hereafter IRDC C from the sample of BT09, from their $\rm N_2D^+$(3-2)
emission observed with {\it ALMA} in Cycle 0 by \citet[][hereafter,
  T13]{2013ApJ...779...96T}, who name the cores C1-N and C1-S. They
are amongst the most promising massive starless/early-stage core candidates to
date. Dynamical study indicates they are moderately sub-virial, unless
a relatively strong, but not exceptional, magnetic field ($\sim$mG) is
present. For this paper, we collected multiple lines of $\rm N_2D^+$
and $\rm N_2H^+$ from a variety of telescopes in order to estimate the
excitation temperatures, column densities and \dfracns.  These results
will then be compared with our chemical models so as to estimate core
ages and constrain dynamical models.

We introduce the observational data in \S\ref{sec:obs} and describe
the measurement of \dfrac in \S\ref{sec:results}. We compare to
chemodynamical models to constrain core ages and collapse rates in
\S\ref{sec:age}.  Discussions and conclusions are presented in
\S\ref{sec:dc} and \S\ref{sec:conc}, respectively.

\section{Observations and Data Reduction}\label{sec:obs}

\begin{figure*}[htb!]
\epsscale{1.}
\plotone{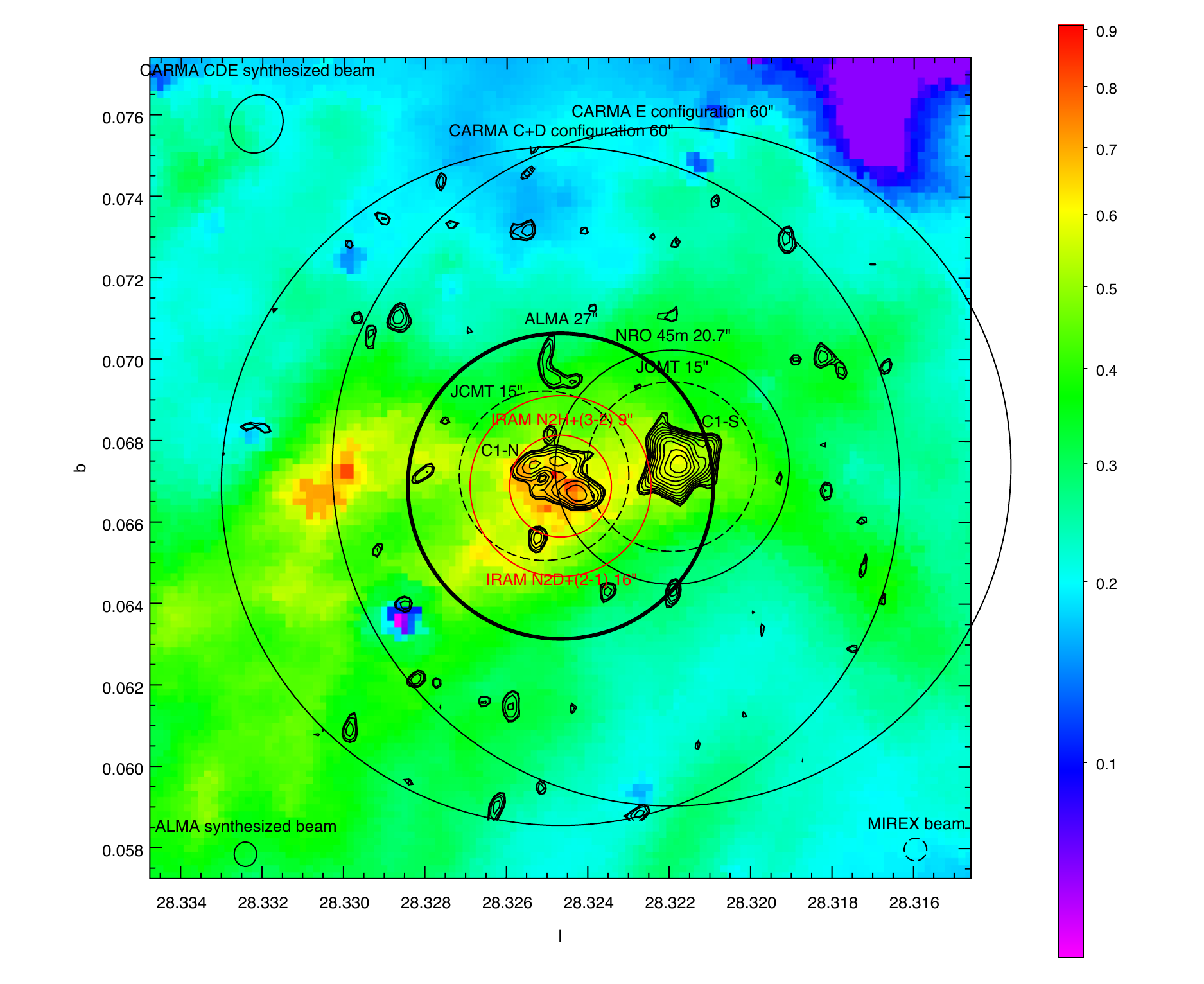}
\caption{
Observation pointings overlaid on the MIREX mass surface density map
\citep{2012ApJ...754....5B}.  The MIREX map is shown in log-scaled color in
unit of g cm$^{-2}$, with the 2\arcsec\ {\it Spitzer} beam size shown
on the lower right corner.  The black contours show \ntdpns(3-2)
integrated intensity from {\it ALMA} Cycle 0 observation \citep{2013ApJ...779...96T},
with the synthesized beam shown on lower left corner.  The circular
shapes represent the primary beams of the various telescopes used in
the observations of multiple transition lines of \ntdp and
\nthpns. Relevant telescopes and transitions are labeled next to the
primary beams, respectively.  The synthesized beam of {\it CARMA} data is
shown on the upper left corner.
\label{fig:pointing}}
\end{figure*}

The two cores were first detected by \ntdpns(3-2) emission in {\it
  ALMA} Cycle 0 observations (T13). The core properties are summarized
in Table \ref{tab:coremodel}. In this paper we use the {\it ALMA} data
from T13 (see their paper for more details of these
observations). Figure \ref{fig:pointing} shows the primary beams of
various observations presented in this paper, along with the two
cores.  Note, C1-S is away from the center of {\it ALMA} primary beam,
so we applied a primary beam efficiency correction (roughly a factor
of 2.0, depending on distance from phase center) to the observed
fluxes (note, this step was not carried out in T13, where the absolute
line fluxes were not utlized in the analysis; note also there is an
error in the normalization of the y-axis of Fig.~4 of T13, which
should be multiplied by a factor of $\simeq0.5$; however, in the case
of C1-S these two corrections effectively cancel each other out; we
note also that the calibration uncertainties of these data are
estimated to be $\lesssim 20\%$).  Those observations whose primary
beam does not fully cover one of the cores will not be used in the
fitting analysis of that core, but we still show the corresponding
spectra for reference. In summary, for C1-N, the valid observations
are: \ntdpns(2-1), \ntdpns(3-2), \nthpns(1-0), \nthpns(3-2),
\nthpns(4-3); for C1-S, the valid observations are: \ntdpns(1-0),
\ntdpns(3-2), \nthpns(1-0), \nthpns(4-3).  Below, we describe the
collection of these data in detail, while Table \ref{tab:obs}
summarizes some important observational parameters.

\begin{deluxetable*}{ccccccccccc}
\tabletypesize{\scriptsize}
\tablecolumns{11}
\tablewidth{0pc}
\tablecaption{Core properties defined by {\it ALMA} observations of \ntdpns(3-2) by T13\tablenotemark{1}\label{tab:coremodel}}
\tablehead{
\colhead{Core} &\colhead{R.A.} & \colhead{DEC.} & \colhead{$\theta_{c}$\tablenotemark{1}} & \colhead{$R_{c}$}&\colhead{$v_{\rm LSR}$} &\colhead{$\sigma_{\rm N_2D^+,obs}$} & \colhead{$\Sigma_{\rm c,mm}$} & \colhead{$N_{\rm H,c,mm}$} & \colhead{$n_{\rm H,c,mm}$}& \colhead{$M_{c,{\rm mm}}$}\\
\colhead{} &\colhead{} & \colhead{} & \colhead{(\arcsec)} & \colhead{(pc)} & \colhead{(\kmsns)} & \colhead{(\kmsns)} & \colhead{(${\rm g\:cm^{-2}}$)} & \colhead{($10^{23}\:{\rm cm^{-2}}$)} & \colhead{($10^5\:{\rm cm}^{-3}$)}& \colhead{$(M_{\odot})$}
}
\startdata
C1-N &$\rm 18^h42^m46\fs89$&$\rm -04\arcdeg04\arcmin06\farcs28$ & 3.38 & 0.0818 & 81.18 & 0.367 & $0.161^{0.321}_{0.0938}$ & $0.688^{1.37}_{0.401}$ & $2.05^{4.12}_{1.10}$ & $16.2^{33.6}_{6.83}$\\
C1-S &$\rm 18^h42^m46\fs50$&$\rm -04\arcdeg04\arcmin15\farcs96$ & 3.61 & 0.0875 & 79.40 & 0.365 & $0.542^{1.08}_{0.322}$ & $2.31^{4.61}_{1.37}$ & $6.43^{12.9}_{3.52}$ & $62.5^{129}_{26.8}$
\enddata
\tablenotetext{1}{
From 4th column, core properties are: Core angular radius; core
physical radius at a distance of 5 kpc; core LSR velocity as defined
by \ntdpns(3-2) emission; observed velocity dispersion of this line
after accouting for hyperfine structure; mean core mass surface
density estimated from 1.3~mm dust continuum emission; equivalent mean
core column density of H nuclei; mean core number density of H
nuclei; core mass. See T13 for further discussion of these physical
properties and their uncertainties.}
\end{deluxetable*}

\begin{deluxetable*}{cccccccc}
\tabletypesize{\scriptsize}
\tablecolumns{8}
\tablewidth{0pc}
\tablecaption{Summary of Observations\label{tab:obs}}
\tablehead{
\colhead{Line} &\colhead{Telescope} & \colhead{Frequency\tablenotemark{4}} &
\colhead{$\theta$\tablenotemark{5}}&\colhead{$\Delta v$\tablenotemark{6}}&\colhead{$\sigma_c$\tablenotemark{7}}&\colhead{$\sigma_s$(C1-N)\tablenotemark{8}}&\colhead{$\sigma_s$(C1-S)\tablenotemark{8}}\\
\colhead{} &\colhead{} & \colhead{(GHz)} &
\colhead{(\arcsec)}&\colhead{(\kmsns)}&\colhead{(mK)}&\colhead{(Jy)}&\colhead{(Jy)}
}
\startdata
\ntdpns(1-0) & {\it NRO 45m} & 77.10924 & 22 & 0.35 & 24 && 0.038(0.7)\\
\ntdpns(2-1) & {\it IRAM 30m}\tablenotemark{1} & 154.21701 & 16 & 0.2 & 17 &0.059(0.4)&\\
\ntdpns(3-2) & {\it ALMA}\tablenotemark{2} & 231.32183 & 2 & 0.08 & 63 &0.046(0.08)&0.082(0.08)\\
\hline
\nthpns(1-0) & {\it CARMA} & 93.17340 & 5 & 0.08 & 290 &0.050(0.32)&0.056(0.32)\\
\nthpns(3-2) & {\it IRAM 30m}\tablenotemark{1} & 279.51176 & 9 & 0.04 & 65 &0.17(0.16)&\\
\nthpns(3-2) & {\it SMA}\tablenotemark{3} & 279.51176 & 4.5 & 0.4 & 78 &0.17(0.8)&0.20(1.6)\\
\nthpns(4-3) & {\it JCMT} & 372.67249 & 15 & 0.2 & 78 &0.75(0.8)&0.49(0.8)\\
\hline
\ohtdpns(1( 1, 0)- 1( 1, 1))& {\it JCMT} & 372.42138 & 15 & 0.2 & 78 &&
\enddata
\tablenotetext{1}{From \citet{2011A&A...529L...7F}}
\tablenotetext{2}{From T13}
\tablenotetext{3}{From Pillai et al., in prep.}
\tablenotetext{4}{http://www.splatalogue.net}
\tablenotetext{5}{Angular resolution}
\tablenotetext{6}{Velocity resolution}
\tablenotetext{7}{Observation rms in \tmbns}
\tablenotetext{8}{Spectra rms in flux density unit after binning (velocity resolution after binning shown in parentheses with unit \kmsns), blank indicates that the data is not used in spectral fitting}
\end{deluxetable*}

\subsection{{\it CARMA}}\label{subsec:obs:carma}

We observed the cores in \nthpns(1-0) at 93~GHz with the {\it CARMA}
15-element array, using the single pointing mode. They were first
observed in D-configuration (beam size $\sim6\arcsec$, October 2012,
bandpass calibrator: 1635+381, phase calibrator: 1743-038, flux
calibrator: Mars) and then in C-configuration (beam size
$\sim3\arcsec$, December 2012, bandpass calibrator: 1635+381, phase
calibrator: 1743-038, flux calibrator: MWC349).  Later, in August 2013
at the {\it CARMA} summer school, they were observed in
E-configuration (beam size $\sim8\arcsec$, bandpass calibrator:
2015+372, phase calibrator: 1743-038, flux calibrator: MWC349).  The
synthesized beam is 5.5\arcsec $\times$ 4.7\arcsec\ with P.A. =
4\arcdeg.  The field-of-view (FOV) is $\sim60\arcsec$, and the largest
detectable scale is $\sim30\arcsec$ (compare to the core sizes
$\la7\arcsec$ in the {\it ALMA} observation).  The spectral resolution
is $\sim0.08$~\kmsns.  The data were reduced with the MIRIAD software
package.  We followed the standard calibration and imaging
procedures. The final 1$\sigma$ rms at the map center for C1 is 0.050
Jy beam$^{-1}$ (combined CDE-configuration). Overall flux calibration
uncertainties are estimate to be $\sim 15\%$.

\subsection{{\it JCMT}}\label{subsec:obs:jcmt}

We observed ortho-H$_2$D$^+$ J$_{k^+,k^-}$ = 1$_{1,0}$ $\rightarrow$
1$_{1,1}$ and \nthpns(4-3) lines towards C1-N and C1-S with the {\it
  JCMT} 15m telescope at 372 GHz (beam size $\sim15\arcsec$).  We used
the HARP instrument, which is a 4$\times$4 receiver array.  Each pair
of adjacent receivers are separated by 30$\arcsec$ and the array has a
total footprint of 2$\arcmin$.  The observation was carried out in the
``jiggle" mode with the Nyquist sampling (1 pointing per 15$\arcsec$).
During our observation some receivers were unavailable so we shifted
the map center so that both cores were well covered in the jiggle
pattern of four adjacent working receivers (H02, H03, H04, H05,
1$\arcmin$ spatial coverage).  We made sure the most massive core C1-S
was at one of the pointing centers.  The system temperature was 286 K.
We adopt a main beam efficiency of 0.64.  The observations were
carried out during the best weather condition at {\it JCMT}, with
$\tau_{225}~<0.05$ and pointing error less than 3\arcsec~on average.
We obtained the calibrated data and used the Starlink software package
to co-add and re-grid the data to construct the cube. Final
sensitivity is shown in Table \ref{tab:obs}. Overall flux
  calibration uncertainties are estimate to be $\lesssim 20\%$.

\subsection{Nobeyama 45m}\label{subsec:obs:nro}

The Nobeyama 45m observations were conducted in May 2013 toward the
C1-S core. We observed \ntdpns(1-0) at 77 GHz, with a beam size of
$\sim$22\arcsec. The data were taken in the position-switching mode.
The TZ receiver was used in combination with the Fast Fourier
Transform Spectrometer (SAM45) providing a bandwidth of 63~MHz and a
frequency resolution of 15.26~kHz (corresponding to 0.05~\kms at the
observing frequency). Pointing was checked by observing the IRC+00363
SiO maser emission every 1 hour, and was shown to be accurate within a
few arc-seconds. The main beam efficiency was 53.4\%.  During the
observation, the system noise temperature was around 170 - 220~K. The
final sensitivity is shown in Table \ref{tab:obs}. Overall flux
  calibration uncertainties are estimate to be $\sim 10\%$.

\subsection{{\it IRAM 30m}}\label{subsec:obs:iram}

The {\it IRAM 30m} data of \ntdpns(2-1) and \nthpns(3-2) presented in
this paper are taken from \citet{2011A&A...529L...7F}. Their spectra
in main-beam temperature have been converted to flux density
(following \S\ref{subsec:texN} equation \ref{eq:TmbtoS}).
  Overall flux calibration uncertainties are estimate to be
  $\sim20\%$. These observations were pointed at C1-N, so we only
include them in the analysis of this core.  However, in the
\ntdpns(2-1) spectrum, there are two velocity components, and one of
them corresponds to the system velocity of C1-S. This is consistent
with the fact that the {\it IRAM 30m} \ntdpns(2-1) observation has a
primary beam partially covering C1-S (see Figure
\ref{fig:pointing}). Given the $\sim$1.8~\kms system velocity
difference between C1-N and C1-S and the good velocity resolution of
the data, we are able to isolate the two cores in velocity space. To
remove the flux contribution from the C1-S component (blue) wing of
the C1-N spectrum, we fit the two velocity components with the CLASS
software\footnote{http://www.iram.fr/IRAMFR/GILDAS}, and subtract the
C1-S velocity component. Again, the {\it IRAM 30m} data are not used
in C1-S analysis.

\subsection{\it SMA}\label{subsec:SMA}

SMA observations were made as part of the ``SMA survey of high-mass
starless cores'' in the most compact configuration (sub-compact) in
two tracks at 279\,GHz in 2007 to 2008. The observations were done in
track sharing mode with multiple sources per track. The correlator was
configured for uniform spectral resolution of $\sim 0.4$\,\kms\ at 279
GHz. Typical system temperatures were between 150 -- 250~K. The gain
calibrators were J$1733-130$, J$1911-201$, and J$1743-038$. The
bandpass calibrator was either 3C273 or 3C454.3, whichever source was
brighter.  The flux calibrators were Uranus, Callisto and Titan.
  Overall flux calibration uncertainties are estimate to be $\sim
  15\%$. The synthesized beam is 5\arcsec$\times$4\arcsec. Further
details on the observing and imaging will be reported in a separate
publication on the survey (Pillai et al. in prep.).

\begin{figure*}[htb!]
\epsscale{1.0}
\plotone{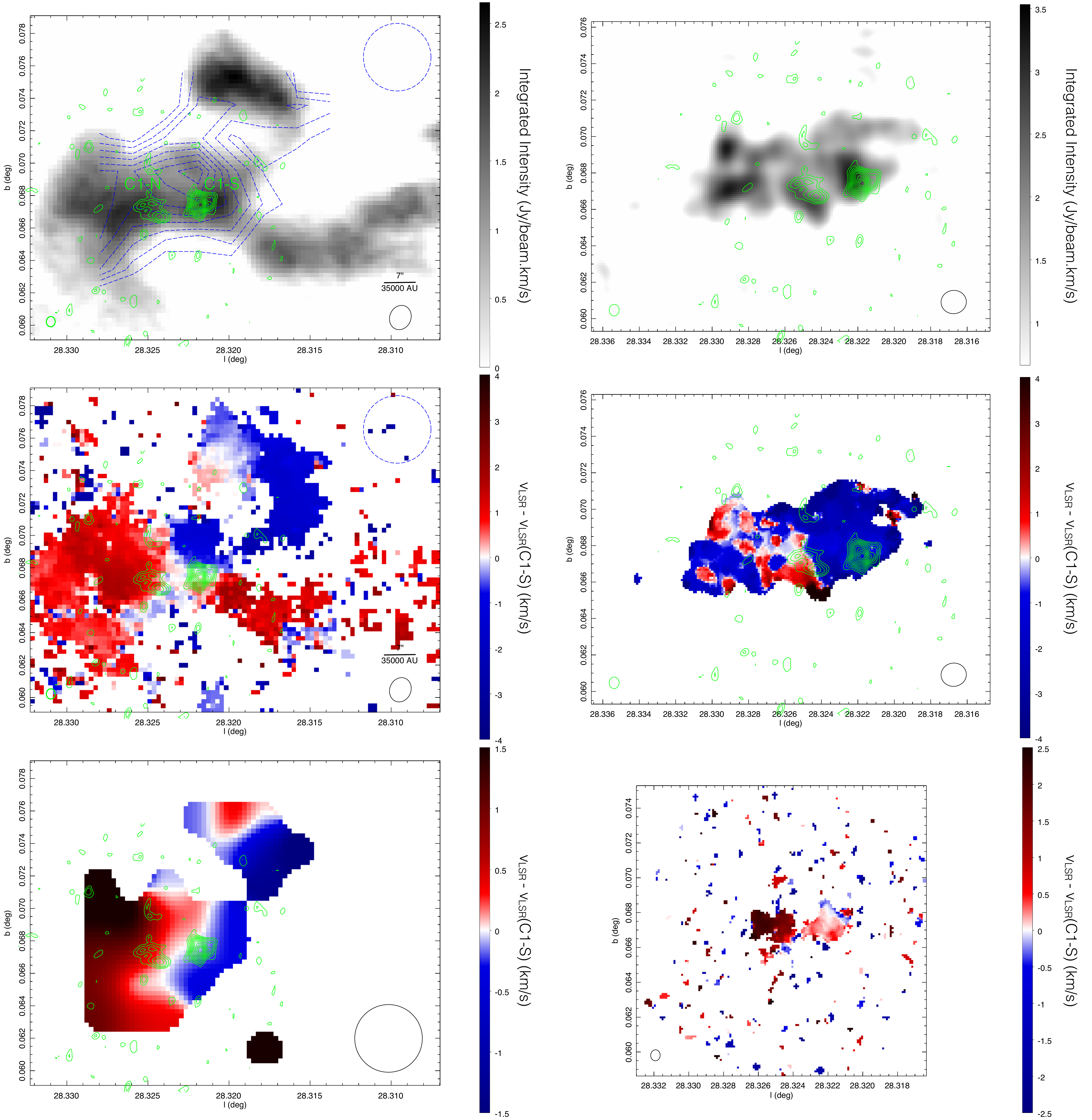}
\caption{
{\it (a) Top left:} Integrated intensities of: \nthpns(1-0) (grey-scale in
units of Jy beam$^{-1}$ \kmsns, with intensities integrated over
velocity range $v_{\rm LSR}=68$ to $90$~\kmsns, i.e., including all
hyperfine structure (HFS); only pixels with SNR $>2$ are shown; noise
at the map center is 0.09~Jy~beam$^{-1}$~\kms and at the map edge is
0.16~Jy~beam$^{-1}$~\kmsns; {\it CARMA} beam is in lower right);
\nthpns(4-3) (blue dashed contours from 3$\sigma$ to 10$\sigma$, with
noise level being $\sigma =$0.10~K~\kmsns, with intensities integrated
over velocity range $v_{\rm LSR}=75$ to $84$~\kms to cover full HFS;
{\it JCMT} beam is shown in upper right); \ntdpns(3-2) showing C1-N and S
cores reported by T13 (green contours from 2, 3, 4 ... 14$\sigma$ with
$\sigma$ = 0.0109 Jy beam$^{-1}$ \kmsns, with intensities integrated
over velocity range $v_{\rm LSR}=76.8$ to $81.9$~\kmsns, covering full
HFS; {\it ALMA} beam is in lower left).
{\it (b) Top right:} Integrated intensities of: \nthpns(3-2)
(grey-scale in units of Jy beam$^{-1}$ \kmsns, with intensities
integrated over velocity range $v_{\rm LSR}=68$ to $90$~\kmsns, i.e.,
including all HFS;  only cells with $>2\sigma$ signal are shown, with
$\sigma=0.33$~Jy~beam$^{-1}$ \kmsns;
{\it SMA} beam is in lower right) and \ntdpns(3-2) (green contours;
same as in (a)).
{\it (c) Middle left:} First moment map of the \nthpns(1-0) isolated
hyperfine component, showing velocities in $\rm km\:s^{-1}$ relative
to $v_{\rm LSR}$(C1-S). {\it CARMA} beam is in lower right. The
\ntdpns(3-2) integrated intensity green contours are shown as in (a),
highlighting the C1-N \& S cores.
{\it (d) Middle right:} First moment map of the \nthpns(3-2) total
HFS, showing velocities in $\rm km\:s^{-1}$ relative to $v_{\rm
  LSR}$(C1-S). {\it SMA} beam is in lower right. The \ntdpns(3-2)
integrated intensity green contours are shown as in (a), highlighting
the C1-N \& S cores.
{\it (e) Bottom left:} First moment map of \nthpns(4-3) emission,
showing velocities in $\rm km\:s^{-1}$ relative to $v_{\rm LSR}$(C1-S)
(integrating over full HFS structure). {\it JCMT} beam is in lower
right. The \ntdpns(3-2) integrated intensity green contours are shown
as in (a), highlighting the C1-N \& S cores.
{\it (f) Bottom right:} First moment map of \ntdpns(3-2) emission,
showing velocities in $\rm km\:s^{-1}$ relative to $v_{\rm LSR}$(C1-S)
(integrating over full HFS structure). {\it ALMA} beam is in lower
left. 
\label{fig:carma}}
\end{figure*}

\section{Results}\label{sec:results}

\subsection{Structure of \nthp Emission Around the \ntdp Cores}\label{subsec:nthpstructure}

Figure~\ref{fig:carma}a shows the integrated intensity imaging of the
C1 region by {\it ALMA} in \ntdpns(3-2), {\it CARMA} in \nthpns(1-0),
and {\it JCMT} in \nthpns(4-3). The {\it ALMA} \ntdpns(3-2) cores are
located within a filament of \nthpns(1-0) emission. However, the {\it
  ALMA} cores appear to be offset from the local {\it CARMA}
\nthpns(1-0) peaks by $\simeq$3.6\arcsec\, corresponding to 18,000~AU
or 0.1~pc at 5~kpc. The map of \nthpns(4-3) shows a peak that is
offset to higher Galactic latitudes from C1-S by $\sim$7.2\arcsec\ (or
0.2~pc). Figure~\ref{fig:carma}b shows the integrated intensity map
from the {\it SMA} observation of \nthpns(3-2). There is a peak of
emission relatively close to the C1-S core, but again offset by about
one core radius. There is a less pronounced concentration of emission
towards C1-N.

These results, especially the \nthpns(4-3) map, suggest there is an
extended envelope of relatively warm gas around the cores. Locally
high volume density is not likely to be the reason for the
\nthpns(4-3) peak, since such a volume density peak should be
associated with a dust continuum peak, which is not apparent in the
1.3~mm emission maps of T13. The \nthpns(4-3) peak does not seem to be
associated with high column density,
as seen in the morphology of the mass surface density map in Figure
\ref{fig:pointing}, where a clear decreasing gradient can be seen from
$b=0.068\degr$ to $b=0.072\degr$.
Rather, it seems more likely that the peak of \nthpns(4-3) emission is
caused by a local volume of gas with higher temperature. Since it is
at the edge of the cloud, it might be caused by moderate shocks from
external gas flows or dissipation of turbulence in the area.
\citet{2006ApJ...651L.125W} reported a water maser detection in this area
(outside C1-S's lowest contour), though at a different velocity
(59.5~\kmsns) and in single channel (0.66 \kmsns).  However, this
water maser was not detected in the more sensitive observations of
\citet{2009ApJS..181..360C}. If it was a real detection, it may be
linked to shock-heated gas in the envelope.  We note that
\citet{2015A&A...577A..75P} have detected CO(8-7) and (9-8) emission
towards the C1-N \& S cores and argue that it is likely that this
emission results from turbulence dissipating in low velocity shocks,
rather than being due to photo-dissociation region (PDR) heating.

Figures \ref{fig:carma}c, d, e and f show the first moment maps
(relative to $v_{\rm LSR}$ of C1-S) of \nthpns(1-0) (isolated
hyperfine component), \nthpns(3-2) (full HFS), \nthpns(4-3) (full HFS)
and \ntdpns(3-2) (full HFS), respectively. The C1-N and S cores are
surrounded by \nthp emitting gas that has broadly the same radial
velocity as that of the \ntdpns(3-2) from the cores, although the
\nthpns(3-2) mean velocity around C1-S is blueshifted by a few km/s,
probably due to the presence of another velocity component, discussed
below. We notice a relatively large velocity gradient ($\sim 1\:{\rm
  km\:s^{-1}} / (2 R_c) \sim 6\:{\rm km\:s^{-1}\:pc^{-1}}$) in
\nthpns(1-0) emission across C1-S. This velocity gradient does not
seem to be influenced by the nearby presence of C1-N, which is located
in a direction that is orthogonal to that of the gradient. The
\nthpns(4-3) emission also shows a gradient across the position of
C1-S (though with much lower resolution), but the direction is
different, and seems likely to be caused by the C1-N to C1-S axis.

In summary, the larger scale kinematics around C1-N and S are
relatively complex and it seems likely that \nthpns(1-0), (3-2) and
(4-3) emission may be dominated by (or at least have significant
contributions from) gas components that are separate from the
\ntdpns(3-2) cores. This will affect our method for estimating the
deuteration fraction in the cores, effectively meaning that we can
only use the \nthpns(1-0), (3-2) and (4-3) spectra extracted from the
core locations to place upper limits on the level of such emission
from the cores. In particular, it is the \nthpns(1-0) data from {\it
  CARMA} and the \nthpns(3-2) data from {\it SMA} that are most
constraining, since they have the most comparable angular resolutions
as the {\it ALMA} observation of
\ntdpns(3-2). Figures~\ref{fig:carma}a and b indicate that
\nthpns(1-0) and (3-2) spectra extracted from the location of the C1-N
and S cores may have $\sim$50\% flux contributions from a larger-scale
clump envelope.

\subsection{Spectra of $\rm N_2D^+$ and $\rm N_2H^+$ Emission Towards the Cores}

With the above considerations in mind, we proceed to analyze the
\nthp and \ntdp spectra extracted from locations of the C1-N and
C1-S cores (i.e., for the {\it CARMA} and {\it SMA} data these are
apertures based on the {\it ALMA} \ntdpns(3-2) core sizes from T13 and
listed in Table~\ref{tab:coremodel}; for the single dish observations,
these are from locations centered on the cores, else as close to the
core positions as allowed). These spectra are shown in
Figure~\ref{fig:C1Nfluxspec} for C1-N and Figure~\ref{fig:C1Sfluxspec}
for C1-S. 

An examination of the spectra indicate that different kinematic
features can be present amongst the different tracers. For example,
the \nthpns(4-3) spectra show different kinematics from the
\ntdpns(3-2) cores.
We perform a hyperfine structure fitting to the \nthpns(4-3) spectrum
in C1-S using the HFS method in CLASS, and the results show that the
velocity width in C1-S is 1.3~\kmsns, much wider than the \ntdpns(3-2)
spectra ($\sim$0.5~\kmsns, see panels c, f in Figure
\ref{fig:C1Sfluxspec} for C1-S). The C1-N \nthpns(4-3) spectrum is
more complicated, since it shows two peaks, with one being at the C1-N
$v_{\rm LSR}$, and the other being at roughly the C1-S $v_{\rm LSR}$.
A fit to the velocity components gives a 0.76~\kms width for C1-N,
moderately larger than the C1-N \ntdpns(3-2) spectra
($\sim$0.5~\kmsns, see panels c, f in Figure \ref{fig:C1Nfluxspec} for
C1-N). These results are also suggestive that the \nthpns(4-3) line in
the C1-S and C1-N regions mostly traces warmer
gas in an envelope external to the cores. The relatively high velocity
dispersion could be caused by shocks. In the following sections we use
the spectra to constrain the excitation temperatures and deuteration
fractions in the cores.

\subsection{Excitation Temperatures, Column Densities and Deuterium Fractions of the \ntdp Cores}\label{subsec:texN}

Here we utilize the \ntdp core models of T13, i.e., two spherical
cores C1-N and C1-S, with the properties listed in Table
\ref{tab:coremodel}. The excitation temperature of \ntdpns, column
densities of \ntdp and \nthp and thus \dfrac of the cores will
be constrained by the multitransition observations of \ntdp and
\nthp in the following way. We construct models that match the
spectra of \ntdpns(3-2), i.e., assuming all of this emission comes
from the cores. These models then make predictions for the other
transitions. For optically thin conditions, the modeled flux from the
cores cannot exceed the observed flux;
however, the modeled flux may be less than that observed if there is a
contribution from a surrounding envelope, either from emission along
the line of sight or from larger angular scales if the cores are
unresolved (i.e., as in the single dish observations).

\begin{figure*}[htb!]
\epsscale{1.}
\plotone{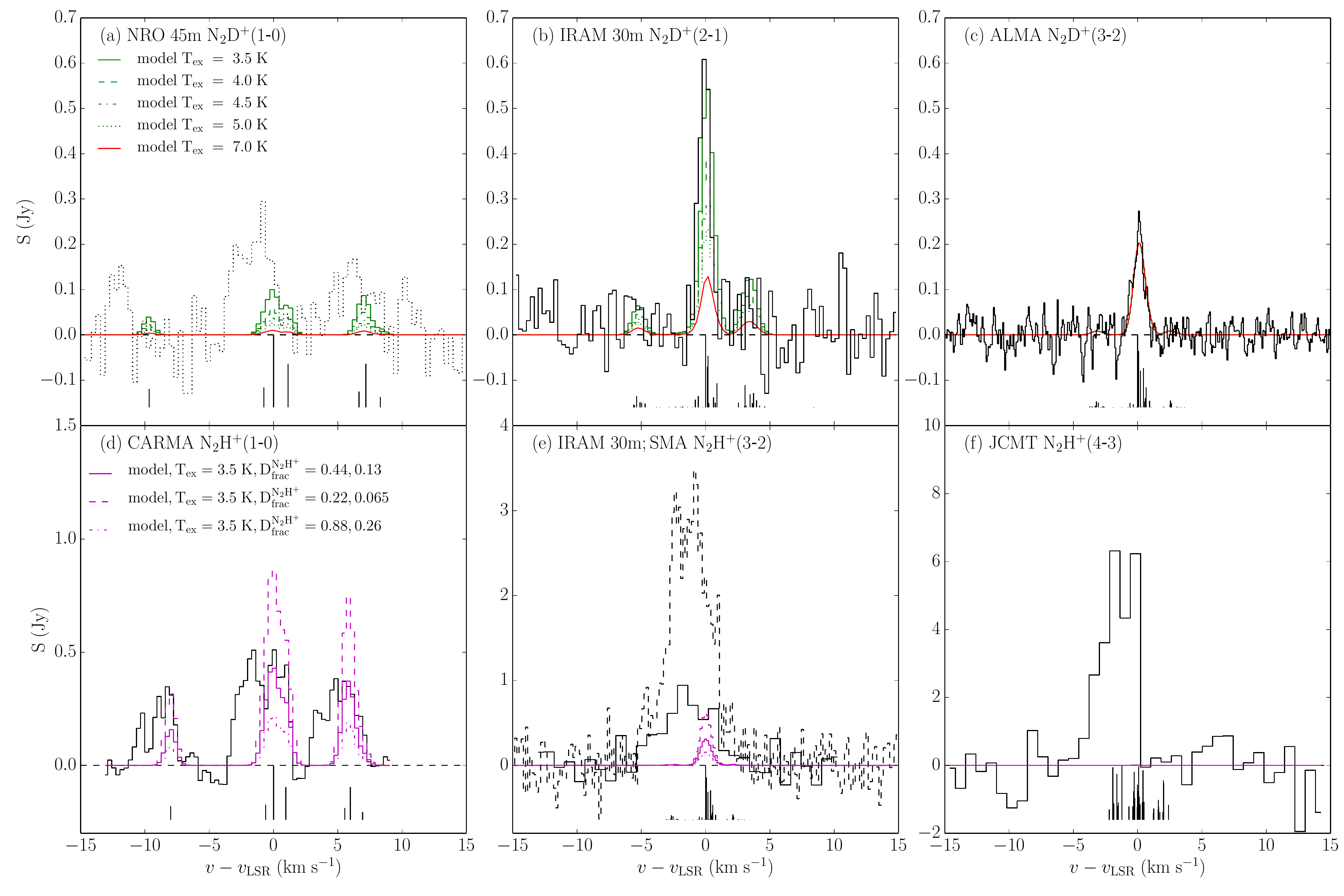}
\caption{
{\it Upper row, panels (a), (b), (c):} Observed \ntdpns (1-0), (2-1),
(3-2) flux density spectra for C1-N (black lines), all shown in the
rest frame of C1-N's $v_{\rm LSR}$ (Table \ref{tab:coremodel}).  The
normalized HFS intensities are shown underneath each spectrum, also in
this velocity frame.
After smoothing, the observed spectra all have peak SNR $>5$. The
resulting spectral resolutions and 1$\sigma$ noise levels are listed
in Table~\ref{tab:obs}.
The model \ntdp spectra, normalized by the {\it ALMA} \ntdpns(3-2)
emission, are shown with green and red lines with various values of
\tex (see legend). Note, the \ntdpns(1-0) data (dotted black line in
panel (a)) is not used for constraining the model because the {\it NRO
  45m} observation was centered on C1-S.  In
\citet{2011A&A...529L...7F}, the \ntdpns(2-1) spectrum has two major
velocity components, with the lower velocity component being -1.8~\kms
away (i.e., from C1-S). We fit hyperfine structures to the spectra and
subtract the C1-S component, leaving the spectrum for C1-N shown in
panel (b).  {\it Lower row, panels (d), (e), (f):} \nthpns (1-0),
(3-2) ({\it SMA} - solid line; {\it IRAM 30m} - dashed line), (4-3)
flux density spectra for C1-N (black lines), again all having peak SNR
$>5$. Modeled \nthp spectra are shown with magenta lines for various
values of \dfrac (see legend, which shows Case 1 and 2 values, see
text). The normalized HFS intensities are shown underneath each spectrum
(note, the \nthpns(4-3) HFS have 60\% flux in the central group).
\label{fig:C1Nfluxspec}}
\end{figure*}

\begin{figure*}[htb!]
\epsscale{1.}
\plotone{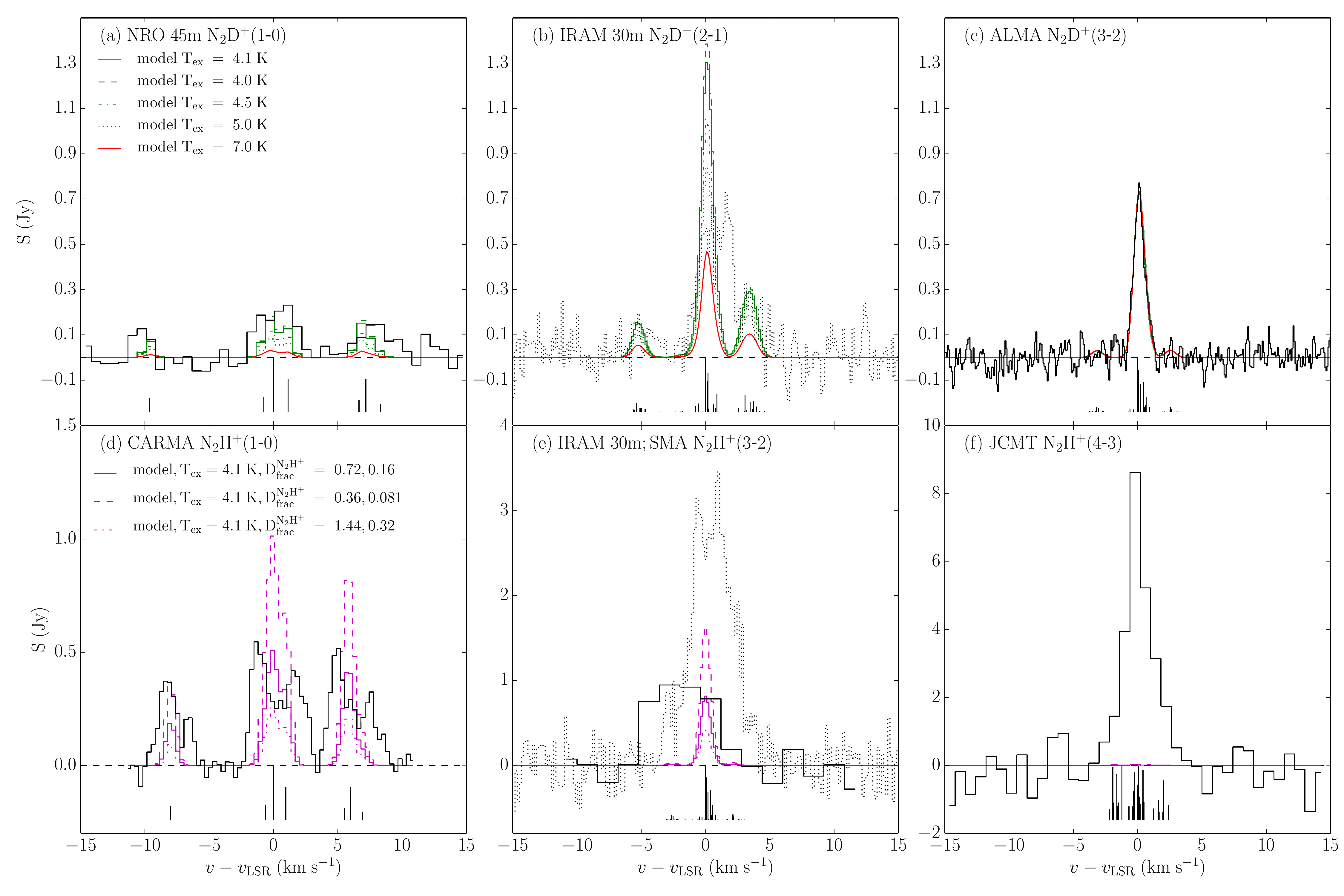}
\caption{ 
Same as Figure \ref{fig:C1Nfluxspec}, but for C1-S. Again, the
smoothed velocity resolution and relevant 1$\sigma$ noise level are
shown in Table~\ref{tab:obs}. {\it IRAM 30m} observations of
\nthpns(3-2) and \ntdpns(2-1), shown with dotted lines, are not used
to constrain the modeling because their pointings were centered on
C1-N.
\label{fig:C1Sfluxspec}}
\end{figure*}

The \ntdpns(1-0), (2-1) and (3-2) and \nthpns(1-0), (3-2) and (4-3)
spectra are shown with black lines in Figure~\ref{fig:C1Nfluxspec} for
C1-N and Figure \ref{fig:C1Sfluxspec} for C1-S.
For the {\it ALMA}, {\it CARMA} and {\it SMA} data, the fluxes are
directly extracted from the cores, since they are resolved. For
single-dish data, spectra in main-beam temperature \tmb are converted
to \snu using
\begin{equation}\label{eq:TmbtoS}
S_\nu = T_{\rm mb} \frac{2k\Omega_{\rm mb}}{\lambda^2},
\end{equation}
where $\Omega_{\rm mb}$ is the main beam solid angle, and $k$ is
Boltzmann's constant. All observed spectra have peak signal-to-noise
ratio (SNR) $>5$, if necessary achieved by smoothing in velocity. The
resulting noise level in each velocity channel is listed in
Table~\ref{tab:obs}.

First we model the \ntdp lines to obtain the best fit for \texns,
which is needed for estimating total column densities. When performing
the multi-transition fitting of the contribution of the C1-N \& S
cores to the \ntdp spectra, we make two assumptions: (1) all
hyperfine components are optically thin (this will be checked by
radiative transfer modeling, below); (2) all hyperfine components have
the same Gaussian profile velocity dispersion, $\sigma_{\rm N_2D^+}$.
All hyperfine components 
are summed to obtain the blended model spectra,
which are to be compared with the observed spectra after normalization
of the integrated intensity and velocity dispersion, which are set by
the {\it ALMA}-observed \ntdpns(3-2), since this defines the cores of
interest.
Note, the {\it ALMA} observation has the best sensitivity, it resolves the
cores, and it filters out emission from large scale structures
($>$9\arcsec).

We vary \tex to reproduce the flux in \ntdpns(2-1) and \ntdpns(1-0)
from the single dish observations as closely as possible, but making
sure the model spectra do not exceed the observed ones. Since these
two \ntdp lines are observed at relatively low angular resolution,
emission from the envelope and/or other larger-scale structures may be
contributing.

Panels (a), (b) and (c) of Figures~\ref{fig:C1Nfluxspec} and
\ref{fig:C1Sfluxspec} show model spectra with a variety of \tex (and
thus a variety of total column densities of \ntdpns) with green and
red lines. Note, since the {\it ALMA} \ntdpns(3-2) line sets the
normalization of equivalent width and velocity dispersion, the green
lines in panels (c) overlap closely with this observed spectrum. Note
also that for \ntdpns(1-0) of C1-N (panel (a) of
Fig. \ref{fig:C1Nfluxspec}) and \ntdpns(2-1) of C1-S (panel (b) of
Fig. \ref{fig:C1Sfluxspec}), we do not have good observed spectra,
since the core is just outside the primary beam. While we display
these spectra here for reference (since they may have some
contribution from the cores), we do not use them to constrain the
model spectra.

The best fit models are shown with solid green lines and the derived
values of \tex and $N_{\rm N_2D^+}$ are listed in Table
\ref{tab:fit}. We derive best-fit excitation temperatures $\sim4\:$K
and column densities of $\sim 6\times 10^{12}\:{\rm cm^{-2}}$ for both
C1-N and C1-S. We refer to these estimates as ``Case 1''. To estimate
the uncertainty caused by noise, we consider a range of models about
the best-fit value that is allowed by the 1$\sigma$ RMS noise of the
spectra, i.e., for its constraint on the height of the peak of the
model spectrum.  These errors are listed in parentheses in Table
\ref{tab:fit}.

The derived values of \tex are $\sim$2~K lower than those adopted by
\citet{2011A&A...529L...7F} (6.4~K), based on hyperfine fitting to
\nthpns(3-2) single-dish observations. As discussed earlier, \nthpns
appears to trace a wider and presumably warmer envelope region
compared to \ntdpns. Deuterated species are likely to trace colder
conditions \citep[e.g.,][]{1999ApJ...523L.165C,
  2013A&A...551A..38P,2015ApJ...804...98K}.
Note, \citet{2005ApJ...619..379C} measured \tex to be about 4.5~K in
a number of low-mass cores, only slightly larger than our derived
values. However, it is also possible that our result of a relatively low
\tex may be explained by the fact that we are fitting two \ntdp
lines, with the lower transition being observed by a single dish
telescope that receives some flux from regions just beyond the
\ntdpns(3-2)-defined cores. 

Our estimates of $T_{\rm ex}$ are relatively low compared to expected
kinetic temperatures of pre-stellar cores, i.e., $\gtrsim 6\:$K
\citep[][for L1544]{2007A&A...470..221C}. The dust temperature in C1-N
\& S is constrained to be $\lesssim 13$~K, from the fact that these
regions appear dark at 70 and even 100~$\rm \mu m$ (T13). At the high
densities of the cores, we would expect gas and dust temperatures to
be reasonably well coupled. Still, subthermal excitation of the
\ntdp lines is a possibility, even though the average volume
densities are close (within a factor of a few) to the critical density
of the \ntdpns(3-2) transition. 

Since there are reasons to expect that our above Case 1 estimates for
$T_{\rm ex}$ may be lower limits due to flux contamination from
extended envelopes, as a ``Case 2'' estimate we will also consider
higher values of \texns.
One possible upper limit is $\sim 10$~K, set by the dust temperature.
However, we note that adopting \texns$ = 10$~K results in a negligible
amount of flux in the \ntdp(1-0) line, which we consider to be
inconsistent with the {\it NRO 45m} observations of C1-S.
\citet{2003A&A...403L..37C} adopted a kinetic temperature of $\sim$7~K
in L1544.  We will use this value of \tex for the Case 2 models,
which are shown by the red lines in Figures~\ref{fig:C1Nfluxspec} and
\ref{fig:C1Sfluxspec}.

To derive the \nthp column density in a core (and thus \dfracns), we
assume this species has the same value of \tex as \ntdpns (for
  the 4~K case, if \nthp has a higher temperature by 1~K, then this
  would increase the estimate of \nthp and \dfracns by 30\%).
However, as shown in \S\ref{subsec:nthpstructure}, \nthp lines show
very extended emission around C1-N and C1-S. In addition, the
temperature of the envelope gas could differ from those in the cores,
likely being higher. Therefore, flux from the \nthp envelope is likely
to be contributing to (and perhaps dominating) the spectra, especially
in the single dish observations of higher $J$ transitions.

Therefore, in fitting the model of core emission to the \nthp spectra
we assume the best fit is achieved when the peak flux density of the
model spectrum reaches the observed flux density, which in practice
will be constrained by the isolated component of the {\it CARMA}
\nthpns(1-0) spectrum and the {\it SMA} \nthpns(3-2) observation.
For the \nthpns(1-0) emission, compared to the main hyperfine
component group at $v-v_{\rm LSR}~\sim~0$~\kmsns, the isolated
component (at negative relative velocity) is more likely to be
optically thin.
Also, given the considerations of \S\ref{subsec:nthpstructure}, we
expect only $\sim 50\%$ of the flux of the observed \nthpns(1-0)
spectra to come from the \ntdp core, with the rest coming from the
clump envelope. However, we will consider a range of 25\% to 100\%,
i.e., a factor of two either side of the central value, as an inherent
uncertainty in this estimate, which will thus translate into a similar
uncertainty in the derived $N_{\rm N_2H^+}$
and \dfracns.  Given the relatively poorer sensitivity of the {\it
  SMA} \nthpns(3-2) observations, we will use this as a consistency
check for the above fitting procedure. We note that these \nthpns(3-2)
spectra show a significant velocity spread to negative velocities
\citep[consistent also with the observations
  of][]{2010ApJ...713L..50C}, which indicates that additional
kinematic components that are separate from the \ntdpns(3-2) cores
could be contributing flux even at the systemic velocities of the
cores.

Panels (d), (e) and (f) of Figures \ref{fig:C1Nfluxspec} and
\ref{fig:C1Sfluxspec} show the observed (black lines) and core model
(magenta lines) \nthp spectra. From the relative heights of the
\nthpns(1-0) hyperfine groups, we see that the main group components
are likely to be affected by optical depth. The best-fit models
(assuming 50\% of the \nthpns(1-0) isolated component comes from the
core) are shown with solid lines, and the relevant derived column
densities of \nthp and thus values of \dfrac are listed in Table
\ref{tab:fit}.

\begin{deluxetable*}{cccccc}
%\tabletypesize{\scriptsize}
\tablecolumns{6}
\tablewidth{0pc}
\tablecaption{Excitation Temperatures, Column Densities and Deuterium Fraction\label{tab:fit}}
\tablehead{
\colhead{Model} &\colhead{Core} &\colhead{\texns\tablenotemark{a}} & \colhead{$N_{\rm N_2D^+}$} & \colhead{$N_{\rm N_2H^+}$\tablenotemark{b}} &
\colhead{\dfracns\tablenotemark{b}}
\\
\colhead{} &\colhead{} &\colhead{(K)} & \colhead{(10$^{13}$cm$^{-2}$)} & \colhead{(10$^{13}$cm$^{-2}$)} &
\colhead{}
}
\startdata
Case 1 & C1-N & 3.50(0.16) & 0.56(0.20) & 0.63 -- 1.26(0.26) -- 2.52 & 0.22 -- 0.44(0.10) -- 0.88\\
... & C1-S & 4.12(0.22) & 0.59(0.21) & 0.41 -- 0.82(0.14) -- 1.64 & 0.36 -- 0.72(0.15) -- 1.4\\
\tableline
Case 2 & C1-N & 7.0 & 0.029 & 0.065 -- 0.13 -- 0.26 & 0.081 -- 0.16 -- 0.32\\
... & C1-S & 7.0 & 0.083 & 0.28 -- 0.55 -- 1.11 & 0.075 -- 0.15 -- 0.30
\enddata
\tablenotetext{a}{Derived from \ntdp fitting}
\tablenotetext{b}{Central values based on fitting to 50\% of observed \nthpns(1-0) isolated component, with error in parentheses based on noise; range set by assuming 25\% to 100\% of this flux (see text)}
\end{deluxetable*}

As a check on the optically thin assumption,
we calculate the optical depth of line emission from the model cores
using
RADEX\footnote{\url{http://www.sron.rug.nl/~vdtak/radex/radex.php}}
\citep{2007A&A...468..627V}.  The common input parameters are kinetic
temperature, \tk=10~K, and line width, $\Delta v$ = 0.9 \kmsns. Then
for C1-N we set H$_2$ number density \nht $= 1.02\times10^5\:{\rm
  cm}^{-3}$ and \nthp column density $N_{\rm N_2H^+} =
2.52\times10^{13}\:$cm$^{-2}$, yielding maximum optical depths for the
\nthpns(1-0) (isolated component), \nthpns(3-2) and \nthpns(4-3) lines
of 0.54, 0.40, 0.016, respectively. Similarly, for C1-S we set \nht
$=3.21\times 10^5\:{\rm cm}^{-3}$ and $N_{\rm N_2H^+} =
1.64\times10^{13}\:$cm$^{-2}$, yielding maximum optical depths 0.21,
0.39, 0.029, respectively, for these same lines. We
expect that \ntdp lines are less affected by self-absorption than
the \nthp lines, given that their column density is a factor of a
few smaller. The estimated optical depths are relatively small, with
the largest effect being for \nthpns(1-0). Given the uncertainties in
core structure that preclude construction of an accurate radiative
transfer analysis, for simplicity we continue with our assumed
optically thin modeling results, but acknowledge that a correction for
\nthpns(1-0) optical depth would lead to smaller estimates of \dfracns
by a factor of about 0.6 for C1-N and 0.8 for C1-S.

As discussed above, for the higher $J$ \nthp lines, the model core
spectra only reproduce small fractions of the total observed flux,
which is likely due to there being a dominant contribution from
larger-scale, warmer envelope gas.
To illustrate the excitation conditions that are needed for the higher
$J$ emission, we calculate the line ratios between \nthpns(3-2) and
\nthpns(4-3) seen in the spectra of Figures~\ref{fig:C1Nfluxspec} and
\ref{fig:C1Sfluxspec}
and compare with results from RADEX models.  The models explore a grid
of physical conditions, with 10$^5$ cm$^{-3}$ $\le$ \nh $\le$ 10$^6$
cm$^{-3}$, 5 K $\le$ \tk $\le$ 30 K, and other fixed parameters,
including $N_{\rm N_2H^+} = 10^{13}\:{\rm cm}^{-2}$,
velocity width = 1.0 \kmsns.
The best-fitting models have \tk $\simeq 28$~K, with the majority of
models requiring \tk $\ga$ 20 K. Since the kinetic temperature \tk is
only at most $\sim$13 K in the C1-N and C1-S cores (T13), this
supports the interpretation that this emission comes from a warmer,
perhaps shock-heated, envelope regions.

\subsection{\ohtdp Abundance}\label{subsec:ohtdp}

\begin{figure*}[htb!]
\epsscale{.57}
\plotone{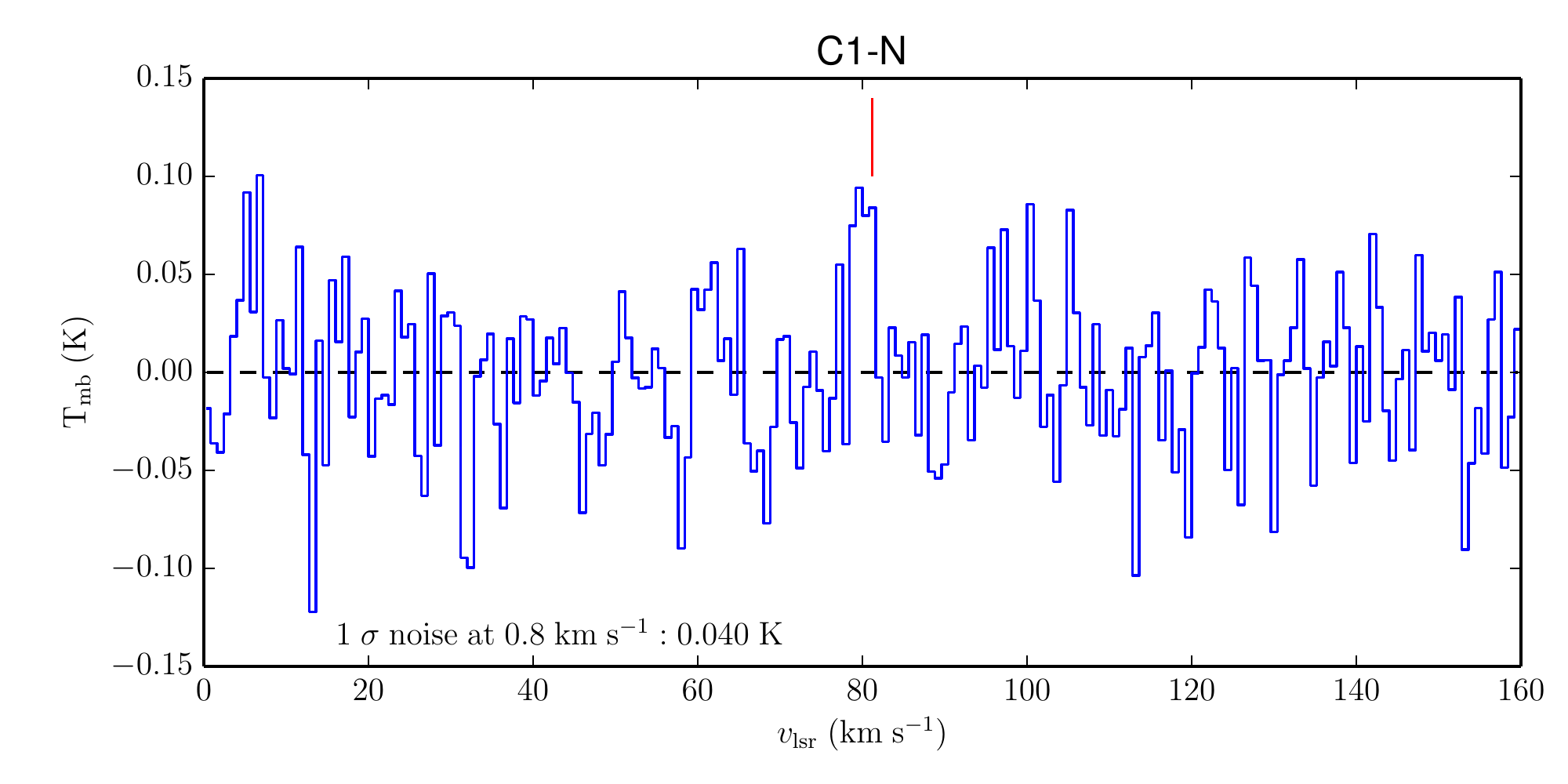}
\plotone{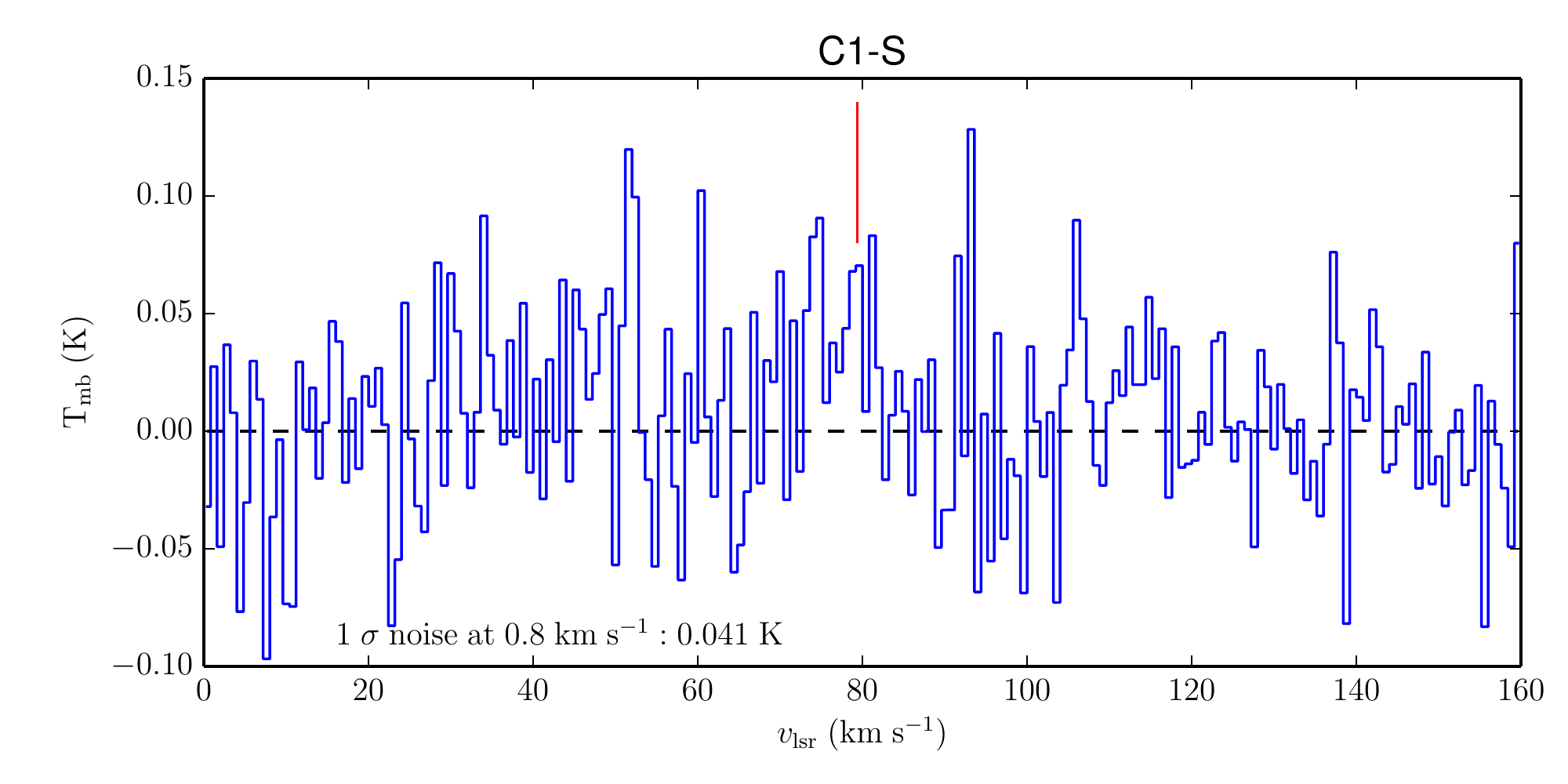}
\caption{
\ohtdp spectra for C1-N and C1-S.  The binned velocity
resolution is 0.8 \kms to potentially maximize SNR, considering the
total width of \ntdp spectra from {\it ALMA} are $\sim$0.9 \kmsns.  The red
vertical lines mark the $v_{\rm LSR}$ of the cores.  RMS noise levels are
labelled in the figure panels.
\label{fig:oh2dp}}
\end{figure*}

Figure \ref{fig:oh2dp} shows the {\it JCMT}-observed \ohtdp spectra
for C1-N and C1-S. To maximize SNR, we binned the spectra to have 0.8
\kms spectral resolution, considering that the line widths of
\ntdpns(3-2) are $\sim$0.9 \kmsns.
%This will maximize the SNR in spectra.  
There is no obvious detection around $v_{\rm LSR}$ of the cores at a
level of 3$\sigma$. However, over a few channels close to these
$v_{\rm LSR}$ values there is a tendency of a lack of negative \tmbns,
which may indicate a tentative detection.
We follow \citet{2003A&A...403L..37C} to calculate the column density.
The \ohtdp excitation temperature is uncertain.
\citet{2003A&A...403L..37C} assumed LTE and adopted \texns=7~K (i.e.,
the value of \tk in L1544). In T13, we estimated \tk~$\lesssim$~10 K
from dust temperature.  However, \tk could be as low as 6~K in some
low-mass cores \citep[e.g.,][]{2007A&A...470..221C}.  Here we adopt a
range of \tex from 4~K (allowing for subthermal excitation) to 10 K
and set \tex = 7 K as a fiducial value. Then we divide the column
density of \ohtdp by $N_{\rm H}$ (estimated from mm continuum in T13,
see Table~\ref{tab:coremodel}) to obtain the abundance of \ohtdpns.
This results in a band of [\ohtdpns] upper limits.  For C1-N, this
band is from 2.4$\times$10$^{-11}$ to 7.3$\times$10$^{-10}$.  For
C1-S, the range is from 0.72$\times$10$^{-11}$ to
2.3$\times$10$^{-10}$.  Note, the values of these upper limits are
uncertain by at least a factor of several, given \tex and $N_{\rm H}$
uncertainties. Later, we will use these results to constrain
astrochemical models.

\citet{2003A&A...403L..37C} measured [\ohtdpns] in L1544 to be
5.5-10$\times$10$^{-11}$, 
depending on the assumption of \tex (note we have expressed
abundances relative to H nuclei, rather than $\rm H_2$).
These values for L1544 happen to be within our estimates of the
$3\sigma$ upper limits in C1-N and C1-S.

\section{Chemodynamical Modeling}\label{sec:age}

We run astrochemical models developed by \citet[][hereafter
  K15]{2015ApJ...804...98K} to compare with the above observational
results. The goal is to obtain the most probable collapse rates 
for C1-N and C1-S.

\subsection{The Fiducial Case}\label{subsec:fid}

The astrochemical models from K15 follow gas phase spin state
chemistry of all relevant 3-atom species along with $\rm H_3O^+$ and
deuterated isotopologues (which are important for O chemistry). K15
also include time-dependent depletion/desorption (TDD) of heavy
elements onto dust grains, starting from some initial assumed
depletion factor, $f_{D,0}$, of heavy elements.

K15 modeled dynamical density evolution (DDE), involving the core
density at time $t$ evolving as
\begin{equation}\label{equ:denschangerate}
\frac{{\rm d}n_{\rm H}(t)}{{\rm d}t} = \alpha_{\rm ff}\frac{n_{\rm H}(t)}{t_{\rm ff}(t)},
\end{equation}
where $t_{\rm ff}$ is the local free-fall time at current density
\nhns, and \aff is a dimensionless parameter setting the collapse
rate. We consider a ``look-back" time, $t_{\rm past}$, relative to the
present time, $t_1$, i.e., related by
\begin{equation}\label{eq:tpast}
t_{\rm past} = t_1 - t.
\end{equation}
So the density at $t_{\rm past}$ is described by
\begin{equation}\label{equ:timedens}
n_{\rm H,past} = n_{\rm H,1}\left[ 1 + 3.60 \alpha_{\rm ff} \left( \frac{n_{\rm H,1}}{10^5\:{\rm cm^{-3}}}\right)^{1/2} \left(\frac{t_{\rm past}}{10^6\:{\rm yr}}\right)\right]^{-2},
\end{equation}
where \nho is the present day density of the core, which we will
define by observation. 

We adopt the core density estimated from dust mm emission in T13,
i.e., \nho$=n_{\rm H,c,mm}$. We have \nhons(C1-N) = 2.0$\times$10$^5$
cm$^{-3}$ and \nhons(C1-S) = 6.4$\times$10$^5$ cm$^{-3}$ (Table
\ref{tab:coremodel}), with uncertainties of about a factor of two.
The other initial conditions and fiducial parameter values are: a
fixed kinetic temperature of 10~K (c.f., the fiducial value of 15~K in
K15); a cosmic ray ionization rate of $2.5\times10^{-17}\:{\rm
  s^{-1}}$; an initial density that is ten times smaller
than the current density, i.e., \nhz = 0.1\nhons;
an initial depletion factor of
C, N, O from the gas phase of \ffdzns~=~3;
and an initial \oprz~=~1.

The fiducial choice of $n_{\rm H,0}/n_{\rm H,1}$, allows exploration
over an order of magnitude change in density, starting from values of
$\sim$few$\times10^4\:{\rm cm^{-3}}$. These initial conditions still
correspond to relatively dense regions of molecular clouds, i.e.,
typical IRDC conditions. Here we expect there to already be moderate
depletion of CO, with \ffd~$\simeq3$ observed by
\citet{2011ApJ...738...11H,2012ApJ...756L..13H}, which thus motivates
the choice of fiducial value. We will also explore models
  with $n_{\rm H,0}/n_{\rm H,1}=0.01$ and \ffdzns~=~1 and 10. The
initial \opr ratio is quite uncertain.  \citet{2011ApJ...729...15C}
measured \opr $\sim0.3-0.8$ in diffuse molecular clouds. We choose
\oprzns~=~1 as a fiducial value, but also explore the effects of
other, especially lower, values, across a range \oprzns~=~0.01-3.
Other parameters of the modeling are the same as those listed in
Tables 1 and 2 of K15.

\begin{figure*}[htb!]
\epsscale{1.}
\plotone{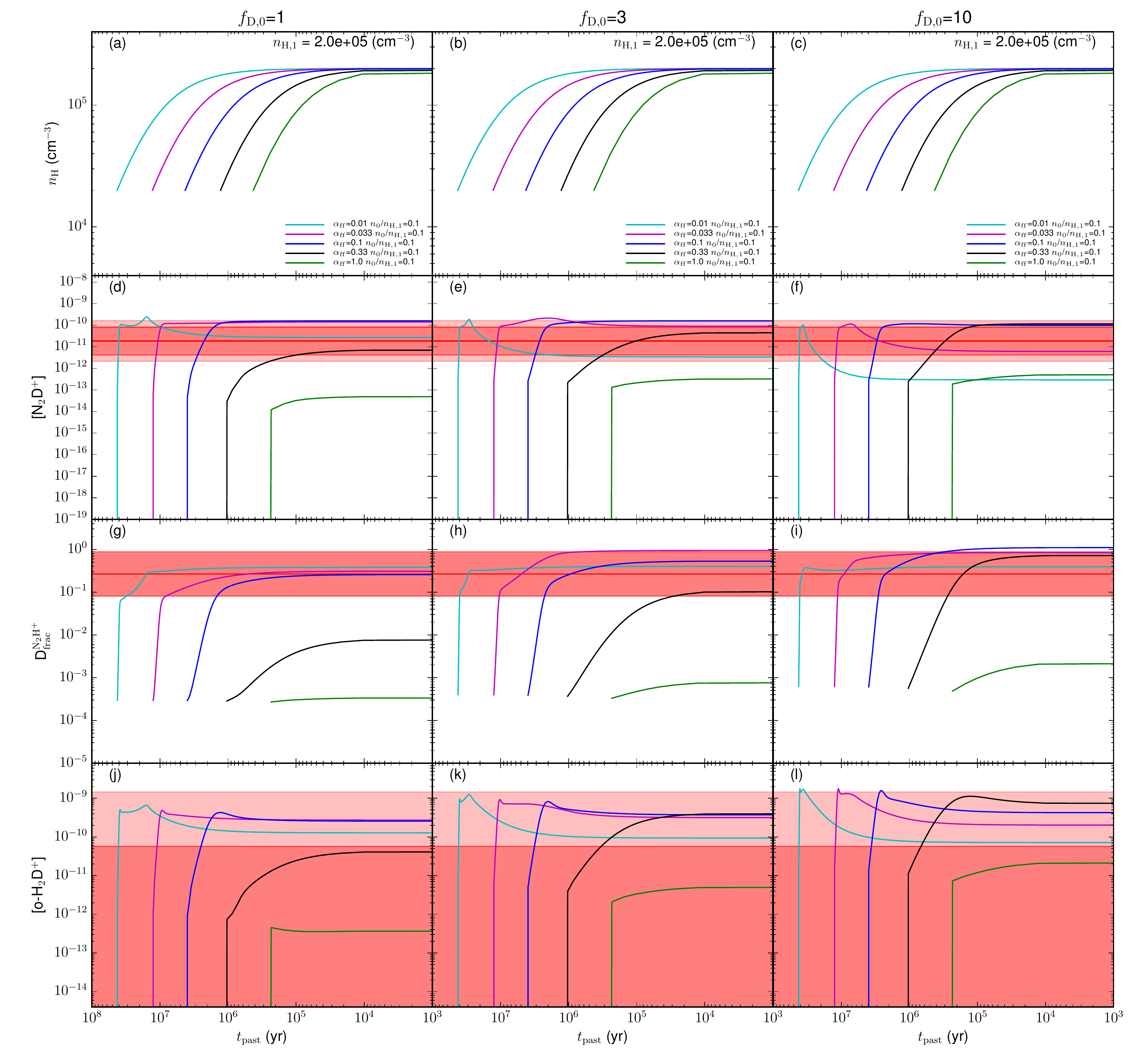}
\caption{
Chemodynamical modeling of C1-N. The models include time dependent
depletion/desorption (TDD) of heavy elements onto dust grains and
dynamical density evolution (DDE),
as parameterized by $\alpha_{\rm ff}$ (see eq. \ref{equ:timedens}).
For C1-N the models have target, present-day density $n_{\rm H,1} =
2.05\times10^5 {\rm cm^{-3}}$. The columns from left to right show
results for initial heavy element depletion factors of \ffdzns= 1, 3
(fiducial), 10. {\it Top row:} Time evolution of density as a function
of $t_{\rm past}$, which increases to the left. Models with
$\alpha_{\rm ff}=0.01,0.033,0.1,0.33,1$ and starting to final density
ratios of $n_{\rm H,0}/n_{\rm H,1}=0.1$ are shown. {\it 2nd row:} Time
evolution of [${\rm N_2D^+}$] for these various models. Case 1 and 2
observational estimates for [${\rm N_2D^+}$] set the darker shaded
region, with additional systematic uncertainties due to $\sim$factor
of two uncertainties in $N_{\rm H}$ shown with a lighter shade.
{\it 3rd row:} Time evolution of $D_{\rm frac}^{\rm N_2H^+}$ for the
same models. The extremes of the Case 1 and 2 estimates for \dfracns
set the range of the shaded region.
Note, the fast collapsing models do not have enough time to reach
large abundances of ${\rm N_2D^+}$ or large values of $D_{\rm
  frac}^{\rm N_2H^+}$.
{(d) Bottom row:} Time evolution of [\ohtdpns] for the same
models. The observational upper limit on [\ohtdpns] is shown with the
red shaded region, with the effect of the \tex uncertainty from 4~K to
7~K shown with a lighter shade (this dominates over the effect of
$N_{\rm H}$ uncertainties).
\label{fig:c1nfiducial}}
\end{figure*}

\begin{figure*}[htb!]
\epsscale{1.}
\plotone{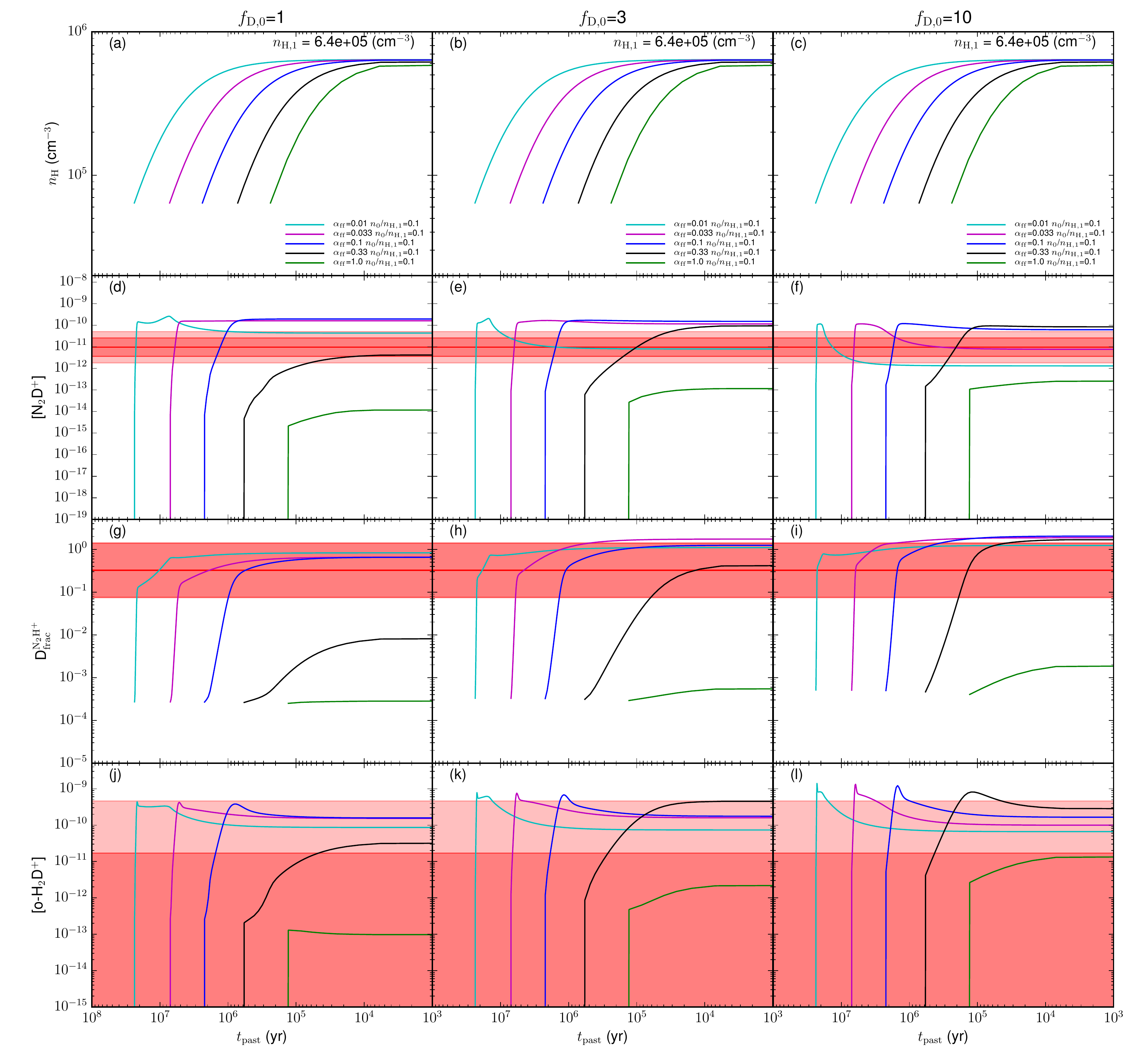}
\caption{
The same as Figure~\ref{fig:c1nfiducial}, but now for C1-S with
target, present-day density $n_{\rm H,1} = 6.43\times10^5 {\rm
  cm^{-3}}$.
\label{fig:c1sfiducial}}
\end{figure*}

Figures \ref{fig:c1nfiducial} and \ref{fig:c1sfiducial} show the
results of the astrochemical modeling of C1-N and C1-S with \ffdzns=1,
3, 10, and including the different rates of density evolution as the
core contracts with $\alpha_{\rm ff}=0.01,0.033,0.1,0.33,1$. The
corresponding evolution of [${\rm N_2D^+}$], $D_{\rm frac}^{\rm
  N_2H^+}$ and [$\rm o$-$\rm H_2D^+$] are also shown. Here square
parentheses denote fractional abundance relative to total H nuclei
number density. The Case 1 and Case 2 observational constraints for
[${\rm N_2D^+}$] and $D_{\rm frac}^{\rm N_2H^+}$, plus the limits on
[$\rm o$-$\rm H_2D^+$] are indicated with the shaded red regions
(additional systematic uncertainties in [\ntdpns] due to $\sim$factor
of two uncertainties in $N_{\rm H}$ are shown with a lighter shade;
the extremes of the Case 1 and 2 estimates for \dfrac define the
shaded region; the effect of the \tex uncertainty from 4~K to 7~K on
the upper limit of [\ohtdpns] is also shown with a lighter shade).

Note that there are also potential systematic uncertainties associated
with the theoretical astrochemical modeling, which for the abundances
and abundance ratios of interest are at approximately the factor of
two level (K15), e.g., as evidenced by the systematic differences of
the results of our chemical network compared to that of
\citet{2015A&A...578A..55S}.

Considering the fiducial \ffdzns~=~3 case for C1-N and C1-S, the
primary effect to note is that in rapidly collapsing cores, i.e.,
$\alpha_{\rm ff}\sim 1$, there is too little time for the level of
deuteration to rise to very high values, so the core exhibits $D_{\rm
  frac}^{\rm N_2H^+} \sim 10^{-3}$. For more slowly evolving cores
with $\alpha_{\rm ff}\lesssim0.3$ there is time for the core to reach
near equilbrium values of [${\rm N_2D^+}$] and $D_{\rm frac}^{\rm
  N_2H^+}\sim 0.1$--1.

\begin{figure*}[htb!]
\epsscale{0.8}
\plotone{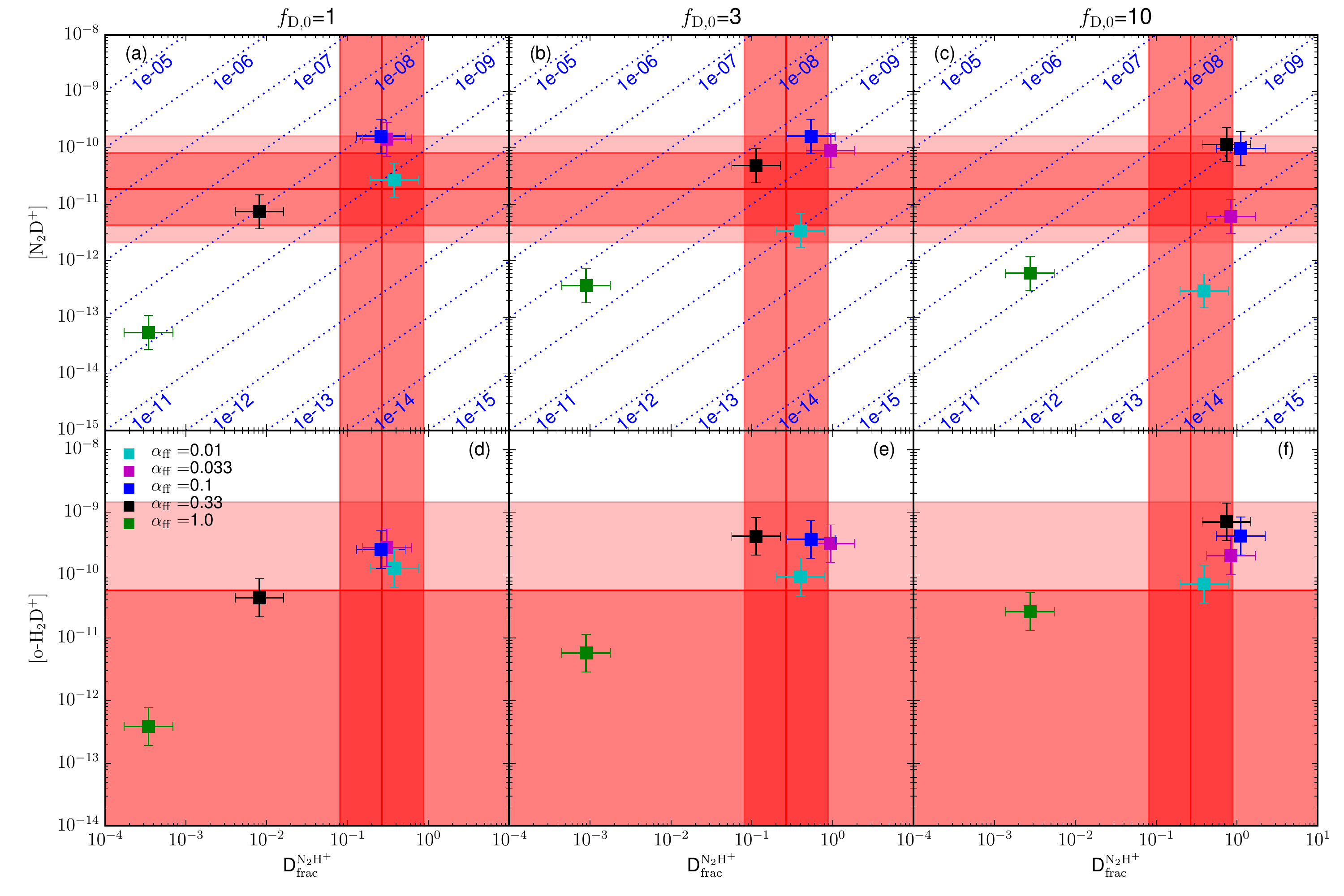}
\caption{
C1-N fiducial models and observational constraints in the [\ntdpns] -
\dfrac (top row) and [\ohtdpns] - \dfrac (bottom row) parameter
space. The blue dotted lines in the top row show constant [\nthpns]
values. The three columns show different initial depletion factors,
$f_{D,0}=1,3,10$, for the fiducial astrochemical models, the results
of which are indicated by the squares, with different colors
representing different collapse rate parameter values (\affns$=0.01$
to 1; see legend). Factor of 2 systematic theoretical errors are
indicated by the error bars around each point. All models are evolved
to the final, observed density starting from a 10 times lower density,
and the initial ortho-to-para ratio of $\rm H_2$ is set to one in all
these cases. The red shaded areas show the same observational
constraints as described in Figure~\ref{fig:c1nfiducial}.
Fast collapsing
\affns$=1$ models are not able to reach the large observed values of
\dfrac (see text).
\label{fig:c1nfancy}}
\end{figure*}

\begin{figure*}[htb!]
\epsscale{0.8}
\plotone{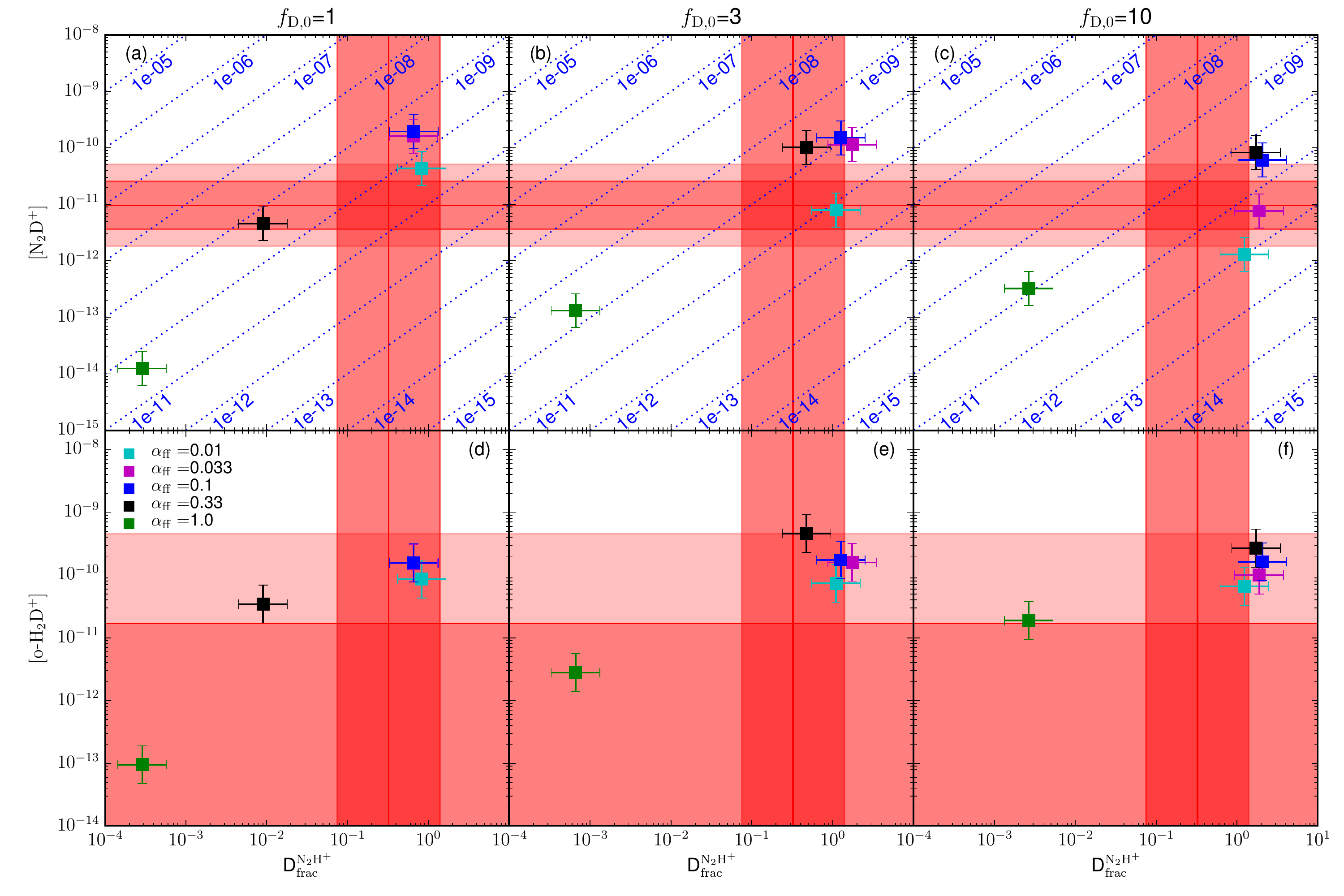}
\caption{
C1-S fiducial results, i.e., the same as Figure~\ref{fig:c1nfancy}, but now for C1-S.
\label{fig:c1sfancy}}
\end{figure*}

Figures \ref{fig:c1nfancy} and \ref{fig:c1sfancy} show summaries of
the modeling results of C1-N and C1-S, respectively. The three
dimensional parameter space of [\ntdpns], \dfrac and [\ohtdpns] is
shown for each of the cases with \ffdzns=1, 3, 10. The locations of
the models at $t=t_1$ (i.e., present-day core conditions) are shown
with the colored square points with factor of two theoretical
uncertainties indicated. Observational constraints on [\ntdpns],
\dfrac and [\ohtdpns] are again depicted by the shaded red regions, as
described above.

In principle, in each panel of Figures \ref{fig:c1nfancy} and
\ref{fig:c1sfancy}, models that fall into the overlapped red areas are
the ones consistent with all the observational constraints, although
leeway should be given for potential theoretical model uncertainties.
The $D_{\rm frac}^{\rm N_2H^+}$ constraints are the most stringent and
the high observed values of $D_{\rm frac}^{\rm N_2H^+} \gtrsim0.1$ for
both C1-N and C1-S allow us to rule out the fastest collapsing
$\alpha_{\rm ff}=1$ models, regardless of the initial depletion factor
(when the core was at a ten times smaller density).
Models with \aff = 0.01, 0.033, 0.1, 0.33 give a much better match to
the observational contraints. In fact, the observational estimates are
broadly consistent with the chemical equilbrium values, which the
slow-collapsing models have time to converge to.

In C1-S some of the slower collapsing models begin to predict
abundances of ${\rm N_2D^+}$ that are moderately higher than the
observational constraints, with the slowest collapsing models with
\aff = 0.01 having the smallest discrepancies. However, these
differences are relatively small (factor of a few), compared to the
difficulties faced by the \aff = 1 models.

\subsection{Parameter Space Exploration}\label{subsec:exploration}

Here we explore the effects of varying model parameters, including
initial ortho-to-para ratio of $\rm H_2$ (\oprzns), cosmic-ray
ionization rate ($\zeta$) and initial density relative to final
density ($\delta n_{\rm H}^\prime\equiv$\nhzns/$n_{\rm H,1}$).  Based
on the results of K15,
temperature variation does not have a significant impact to deuterium
chemistry at $T_k\la$ 15 K.  Since the temperatures in C1-N and C1-S are
$\lesssim13$~K (T13),
we keep the fiducial value
of $T_k=10$~K.

\begin{figure*}[htb!]
\epsscale{0.8}
\plotone{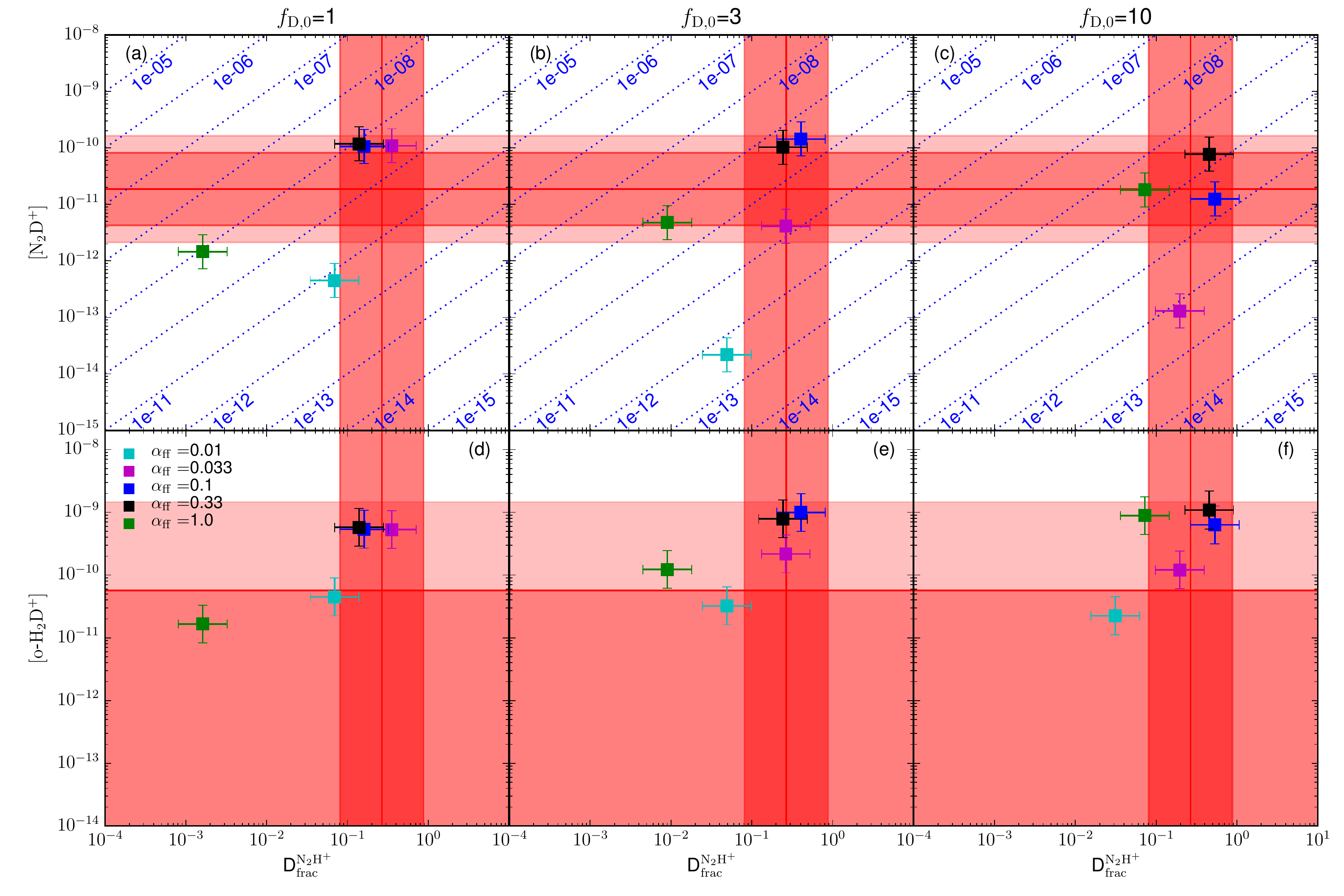}
\caption{
C1-N with high cosmic ray ionization rate, i.e., the same as
Figure~\ref{fig:c1nfancy}, but now the astrochemical models are run
with a higher cosmic-ray ionization rate $\zeta = 10^{-16}\:{\rm
  s}^{-1}$.
\label{fig:c1nfancyhighCR}}
\end{figure*}

\begin{figure*}[htb!]
\epsscale{0.8}
\plotone{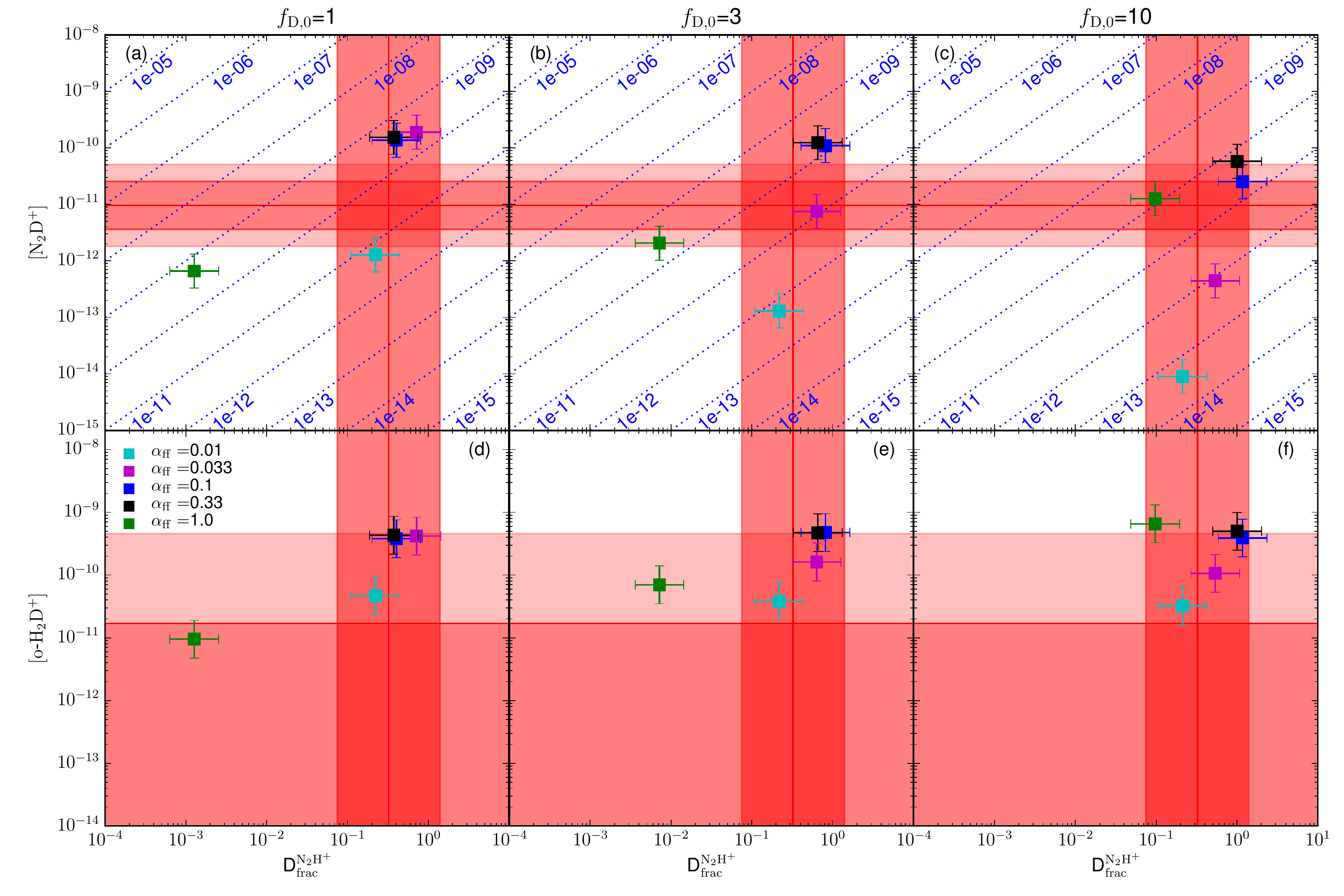}
\caption{
C1-S with high cosmic ray ionization rate, i.e., the same as
Figure~\ref{fig:c1nfancyhighCR}, but now for C1-S.
\label{fig:c1sfancyhighCR}}
\end{figure*}

Figures \ref{fig:c1nfancyhighCR} and \ref{fig:c1sfancyhighCR} show
exploration with the higher cosmic-ray ionization rate $\zeta$ =
10$^{-16}$ s$^{-1}$ ($4\times$ higher than the fiducial value).
Compared to the fiducial models (Figures \ref{fig:c1nfancy} and
\ref{fig:c1sfancy}), there are two notable changes.  First, the fast
collapsing models (\aff = 0.33, 1.0) have higher \dfracns.  In
particular, the \aff = 0.33 model reaches the observed \dfracns, even
if there is no initial depletion. This higher rate of increase of
deuteration 
is a direct consequence of the higher value of $\zeta$.
Consequently, \dfrac in the fast collapsing models is closer to the
equilibrium value. A second change is a decrease in \dfrac of the
slower collapsing models, e.g., \aff$=0.03,0.01$.
This is also expected from the results of K15, where high $\zeta$
reduces the equilibrium \dfracns. The slow collapsing models have more
than enough time to reach this equilbrium value.

\begin{figure*}[htb!]
\epsscale{0.8}
\plotone{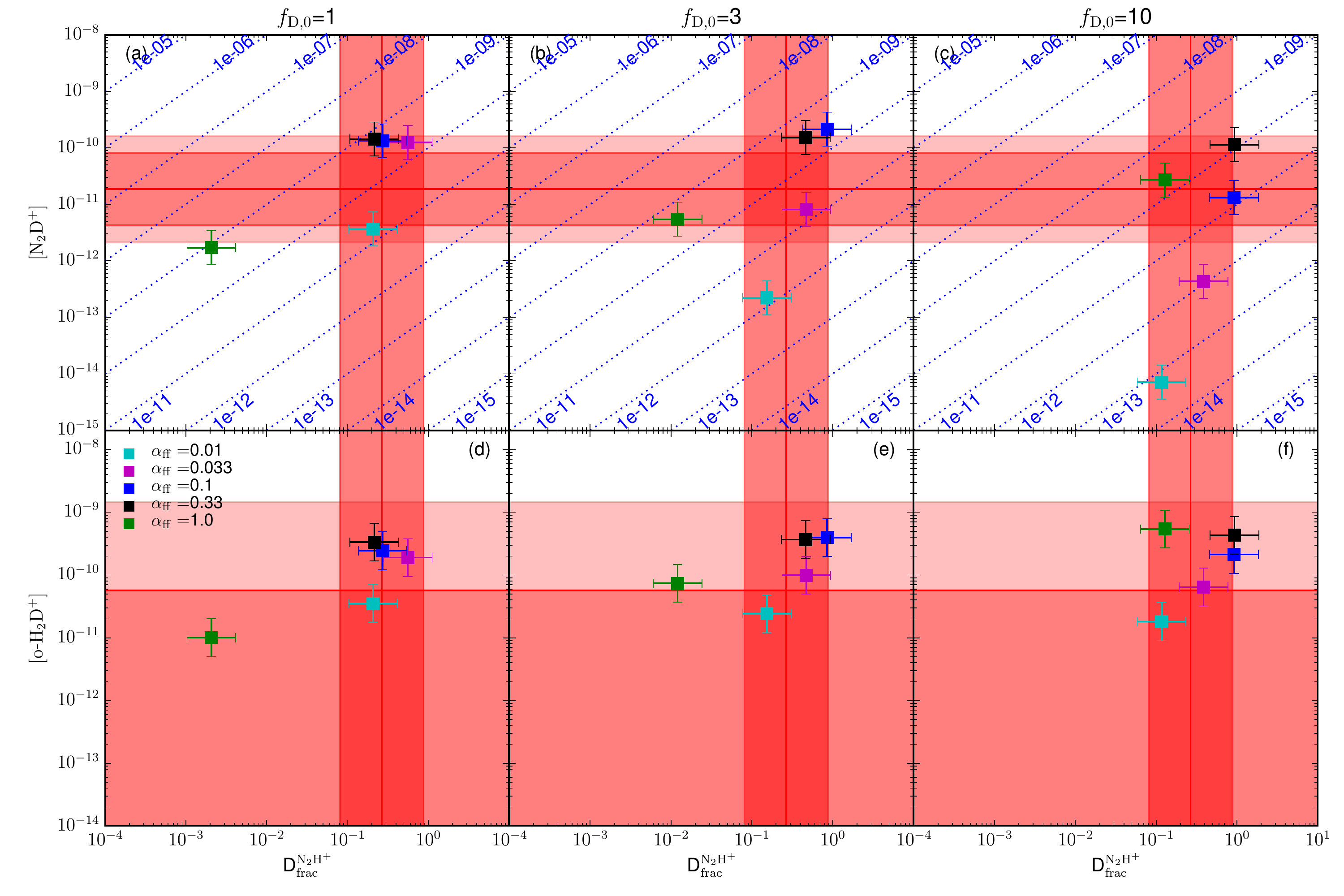}
\caption{
C1-N with low initial density, i.e., the same as Figure
\ref{fig:c1nfancy}, but with \nhz = 0.01\nhons.
\label{fig:c1nfancylowdens}}
\end{figure*}

\begin{figure*}[htb!]
\epsscale{0.8}
\plotone{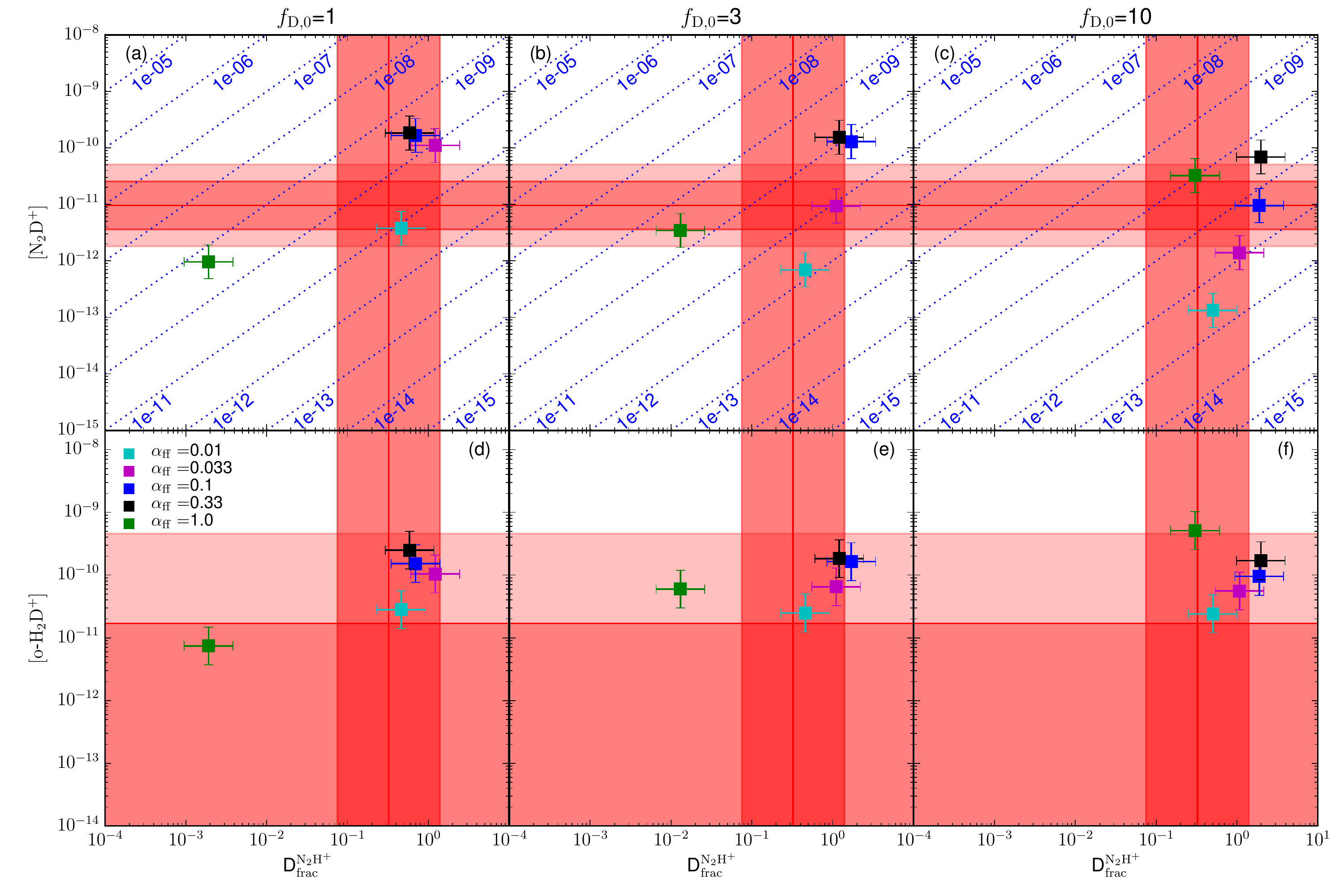}
\caption{
C1-S with low initial density, i.e., the same as Figure
\ref{fig:c1nfancylowdens}, but for C1-S.
\label{fig:c1sfancylowdens}}
\end{figure*}

Figures \ref{fig:c1nfancylowdens} and \ref{fig:c1sfancylowdens} show
the effect of varying the ratio of the initial model density compared
to the final density. The models here have \nhz = 0.01\nho (while
the fiducial case assumed \nhz = 0.1\nhons). 
The main effect of starting with a lower density is that there is more
time for chemical evolution of the gas
so that deuteration equilibrium can be reached in 
faster collapsing models.

\begin{figure*}[htb!]
\epsscale{0.8}
\plotone{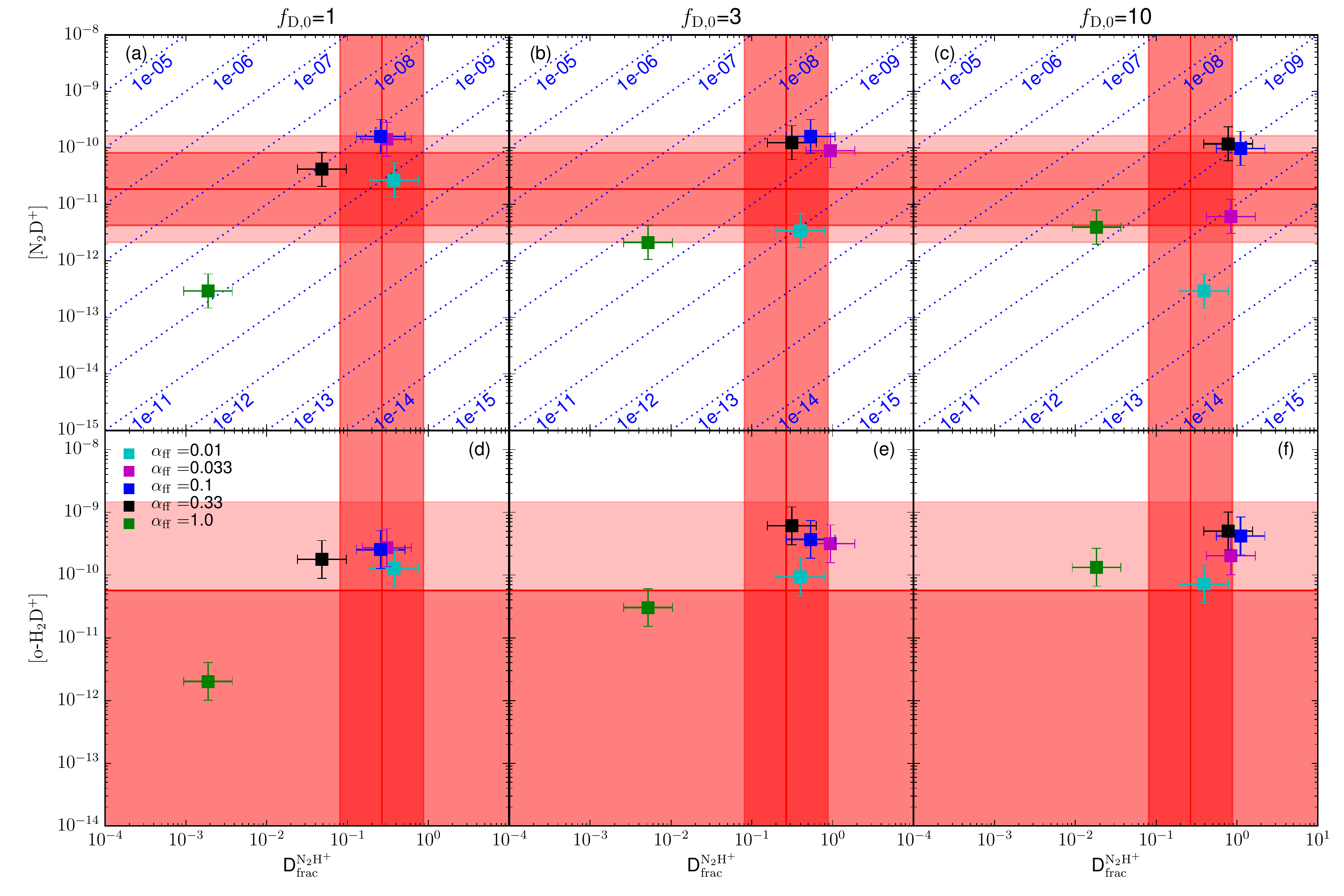}
\caption{
C1-N with low initial ortho-to-para ratio of $\rm H_2$, i.e., the same as Figure
\ref{fig:c1nfancy}, but with \oprz = 0.1.
\label{fig:c1nfancylowopr}}
\end{figure*}

\begin{figure*}[htb!]
\epsscale{0.8}
\plotone{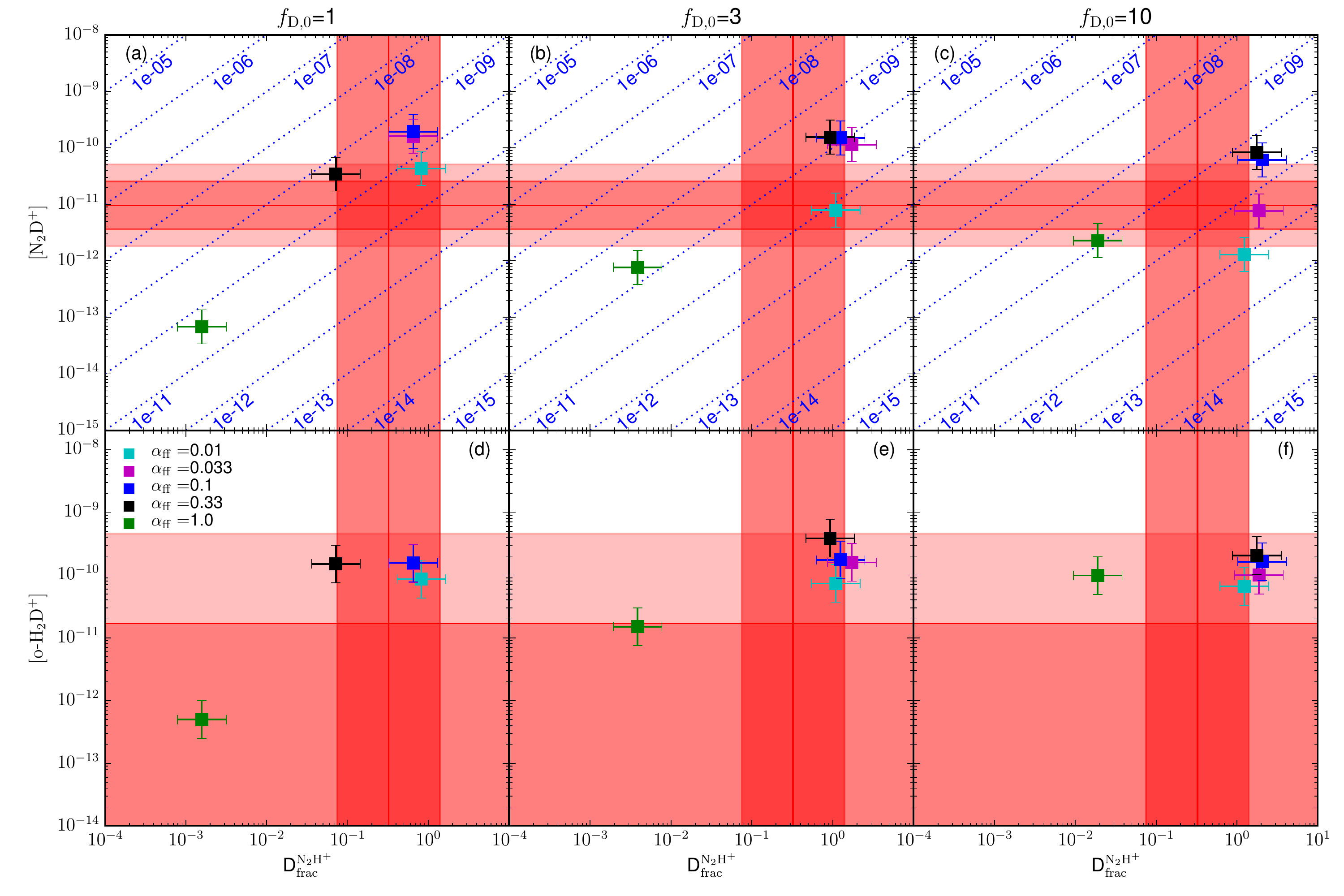}
\caption{
C1-S with low initial ortho-to-para ratio of $\rm H_2$, i.e., the same
as Figure~\ref{fig:c1nfancylowopr}, but for C1-S.
\label{fig:c1sfancylowopr}}
\end{figure*}

Figures \ref{fig:c1nfancylowopr} and \ref{fig:c1sfancylowopr} show
another variation from the fiducial case, with \oprz = 0.1 (compared
to \oprz = 1 in the fiducial models). In general, high \opr
suppresses deuteration. \dfrac does not reach equilibrium until the
ortho-to-para ratio of $\rm H_2$ has dropped significantly.
The typical timescale for ortho-to-para H$_2$ conversion is of order
one Myr, depending on physical conditions, and so is the deuterium
fractionation timescale.
If models start with lower \oprzns, then the establishment of \opr
equilibrium is quicker.  So the main difference between the models
shown here and the fiducial models are that fast collapsing models can
reach higher \dfracns. This helps make the \aff = 0.3 model more
consistent with the observations in terms of \dfracns. 

However, it is also important to note that by lowering the \oprz to be
0.1 for the initial condition, we are in effect starting with a
chemically evolved and therefore relatively old molecular cloud as the
initial condition for dense gas core formation.

\subsection{Best Fit \aff}\label{subsec:bestaff}

For each \affns, we explore the other model parameters, compare with
observational constraints and combine the results to estimate a
likelihood parameter.
The explored parameters are: cosmic-ray ionization rate $\zeta$ =
1.0$\times$10$^{-18}$ s$^{-1}$, 3.3$\times$10$^{-18}$ s$^{-1}$,
1.0$\times$10$^{-17}$ s$^{-1}$, 3.3$\times$10$^{-17}$ s$^{-1}$,
1.0$\times$10$^{-16}$ s$^{-1}$ \citep[extension to lower values allows for
the possibility of magnetic mirror shielding and attenuation of cosmic
rays in dense, magnetized cloud cores][]{2011A&A...530A.109P},
initial density relative to final density $\delta n_{\rm
  H}^\prime\equiv$~\nhzns/\nho $=0.1, 0.01$, initial depletion factor
\ffdz = 1, 3, 10, and initial ortho-to-para H$_2$ ratio \oprz = 3, 1,
0.1, 0.01. For each specific model [\affns, $\zeta$, $\delta n_{\rm
    H}^\prime$, \ffdzns, \oprzns], we calculate its total (summed in
quadrature) ``distance,'' $\Delta$, in the three dimensional log-scale
parameter space to the ``observed location'' of [\ntdpns], \dfracns,
[o-H$_2$D$^+$] normalized by the log-space width of the observational
constraint. The observed location for [\ntdpns] and \dfrac is defined
as the geometric mean value of the upper and lower limits (combining
Cases 1 and 2).  If the model result is between the lower and upper
limits, its contribution to the total distance is set to zero. For
[\ohtdpns], the observed location is set at the upper limit resulting
from \tex = 7 K. If the model value is below [\ohtdpns] at \tex = 7 K,
the distance contribution is zero.  Otherwise it is the log-space
difference (to the 7 K location) normalized by the distance from 7 K
location to the upper limit.  Note we also allow for a potential
factor of two systematic uncertainty in the abundances [\ntdpns] and
[\ohtdpns] (to either higher or lower values) due to the uncertainty
in the observed H column density.

Then, considering the two values of $\delta n_{\rm H}^\prime=0.01, 0.1$ separately, for each \affns, we
average the total distances from each specific model [$\zeta$,
  \ffdzns, \oprzns] to have a likelihood parameter
$\bar{\Delta}(\alpha_{\rm ff},\delta n_{\rm H})$.
Lower $\bar{\Delta}$ means better agreement.  The results are listed
in Table \ref{tab:rhomodel}.  Note, for this simple test, we do not
give special weighting to any of the parameters, i.e., we assume that
all parameters are equally important (which has guided the range of
parameters considered).

\begin{deluxetable*}{ccccccc}
\tabletypesize{\scriptsize}
\tablecolumns{7}
\tablewidth{0pc}
\tablecaption{Likelihood parameters, $\bar{\Delta}(\alpha_{\rm ff},\delta n_{\rm H}^\prime)$, for explored astrochemical models\label{tab:rhomodel}}
\tablehead{
\colhead{Core} &\colhead{$\delta n_{\rm H}^\prime\equiv$~\nhzns/\nhons} & \colhead{\affns=0.01} & \colhead{\affns=0.033} & \colhead{\affns=0.1}&\colhead{\affns=0.33}&\colhead{\affns=1.0}
}
\startdata
C1-S &0.1  &0.99&1.25&1.58&2.49&3.62\\
C1-S &0.01 &1.21&1.00&1.20&1.71&2.70\\
\hline
C1-N &0.1  &1.17&0.96&1.28&2.46&4.05\\
C1-N &0.01 &1.30&0.94&0.86&1.55&2.85
\enddata
\end{deluxetable*}

For both C1-N and C1-S, the best-fitting values of \aff are $\ll1$,
which would suggest that both C1-N and C1-S are contracting very
slowly compared to free-fall colapse.  However, the more meaningful
constraint is that \aff~$\gtrsim0.3$ models are disfavored with their
values of $\bar{\Delta}$ greater than the best-fit models by $\gtrsim
50\%$.

\begin{figure*}[htb!]
\epsscale{1.}
\plotone{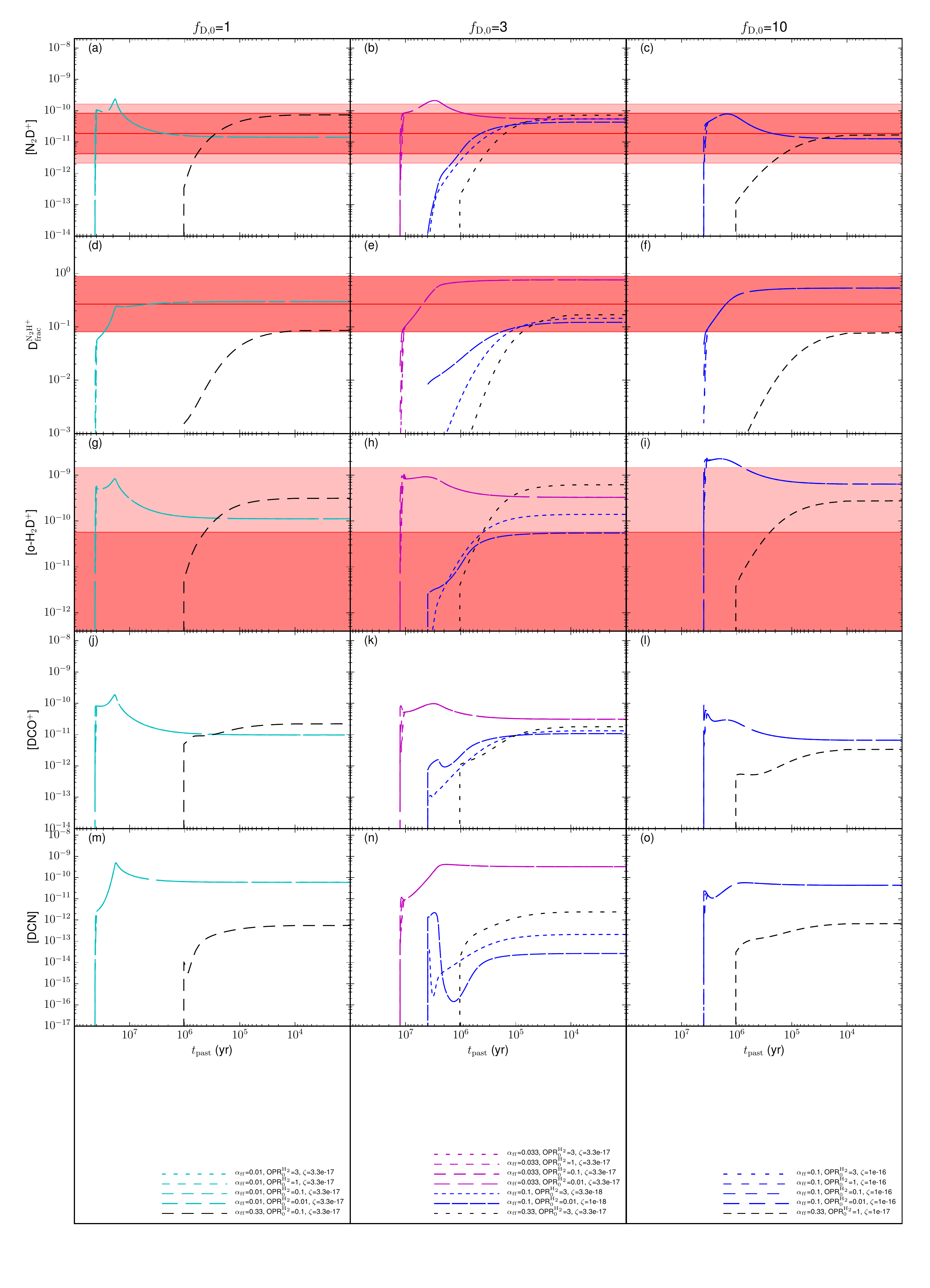}
\caption{
Most promising models for C1-N with $\delta n_{\rm H}^\prime=0.1$ from
the parameter space exploration described in
\S\ref{subsec:exploration}, i.e., models that have final values of
      [\ntdpns], \dfracns, [\ohtdpns] within observational limits. The
      upper three rows follow the same format as the equivalent rows
      in Fig.~\ref{fig:c1nfiducial}. The next rows show time evolution
      of abundances [$\rm DCO^+$] and [DCN], which with future
      observational constraints may help to discriminate between the
      models.
\label{fig:c1ngood10}}
\end{figure*}

\begin{figure*}[htb!]
\epsscale{1.}
\plotone{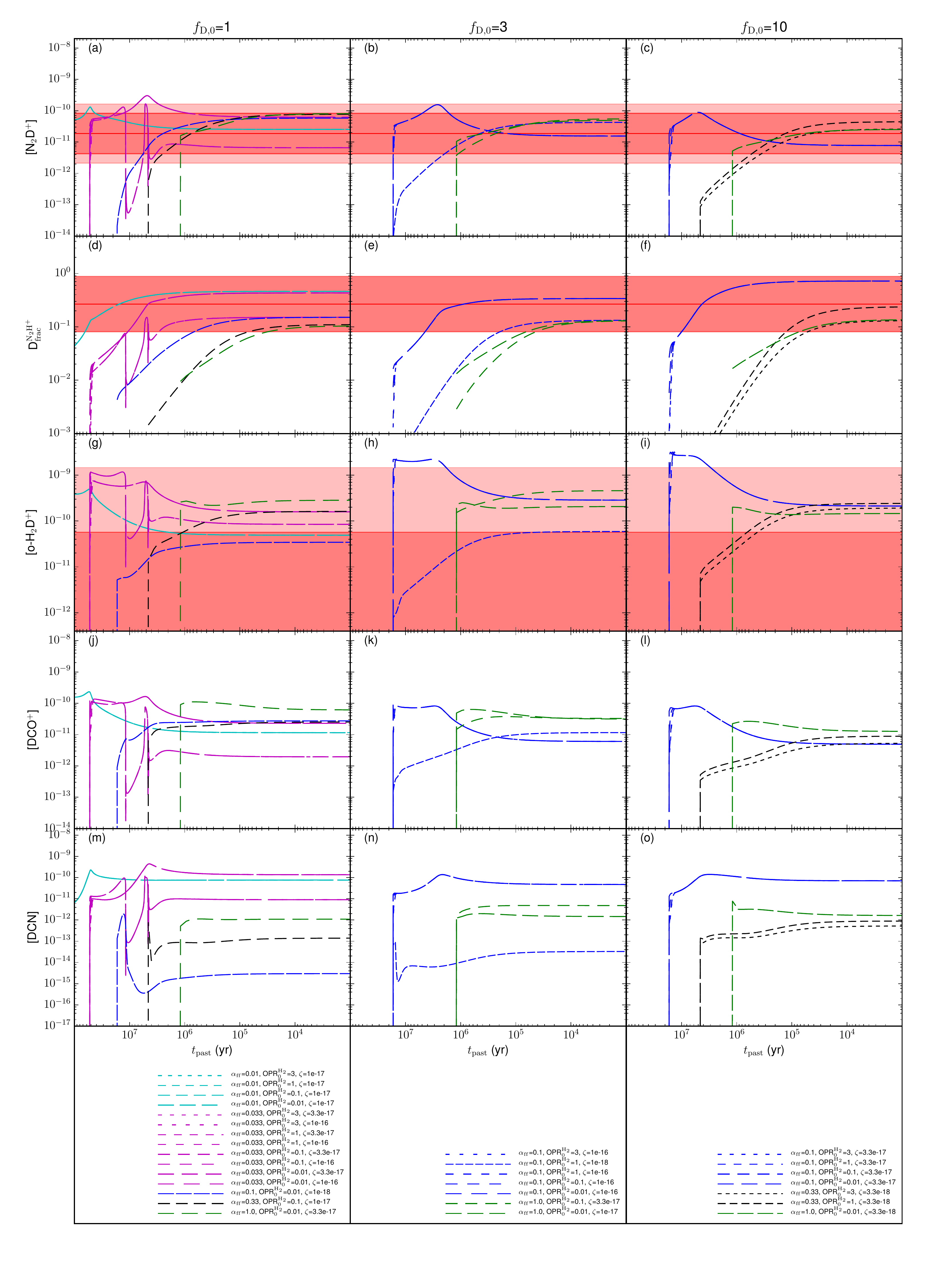}
\caption{
Most promising models for C1-N with $\delta n_{\rm H}^\prime=0.01$, following format of Fig.~\ref{fig:c1ngood10}.
\label{fig:c1ngood100}}
\end{figure*}

\begin{figure*}[htb!]
\epsscale{1.}
\plotone{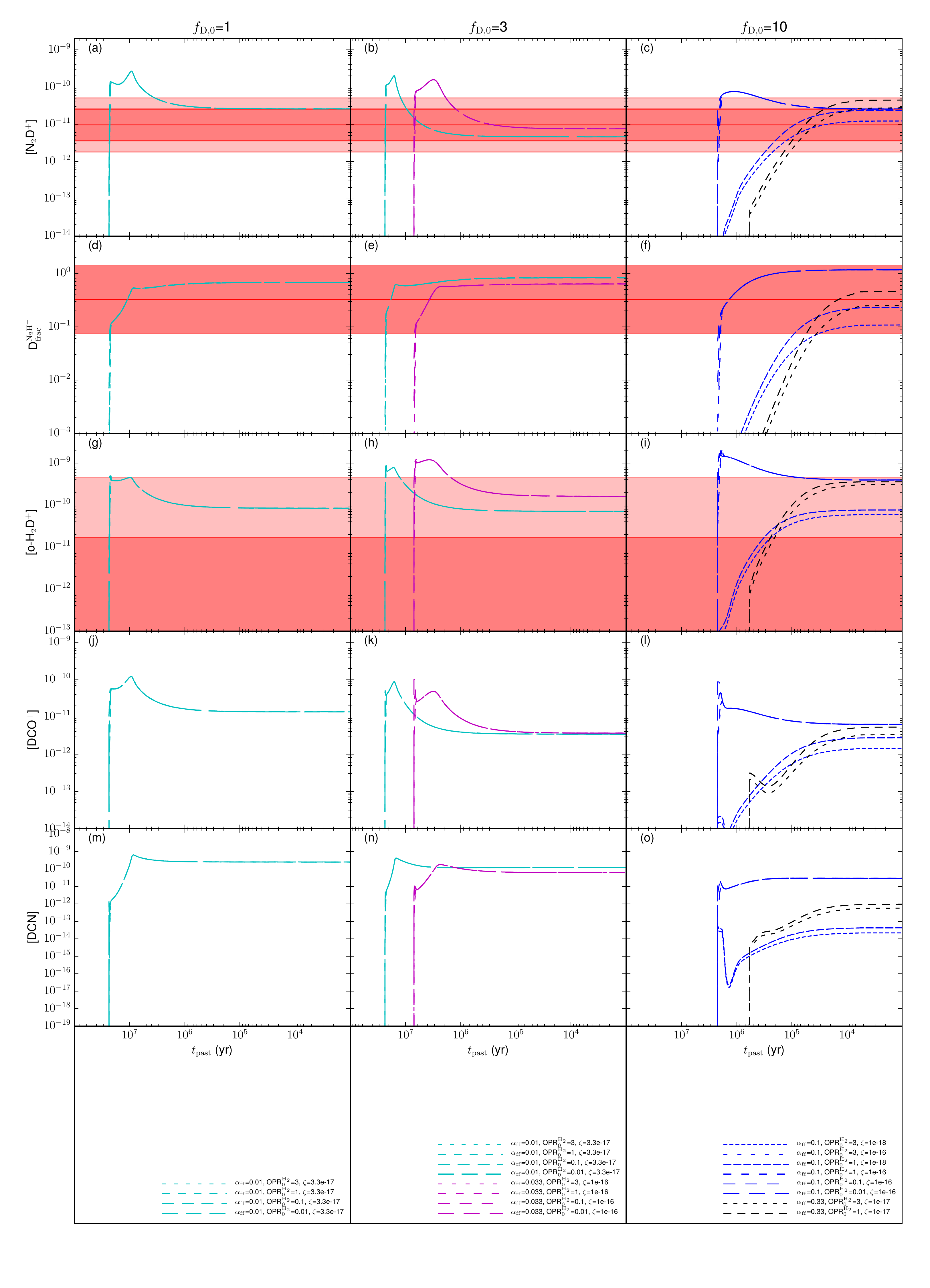}
\caption{
Most promising models for C1-S with $\delta n_{\rm H}^\prime=0.1$, following format of Fig.~\ref{fig:c1ngood10}. 
\label{fig:c1sgood10}}
\end{figure*}

\begin{figure*}[htb!]
\epsscale{1.}
\plotone{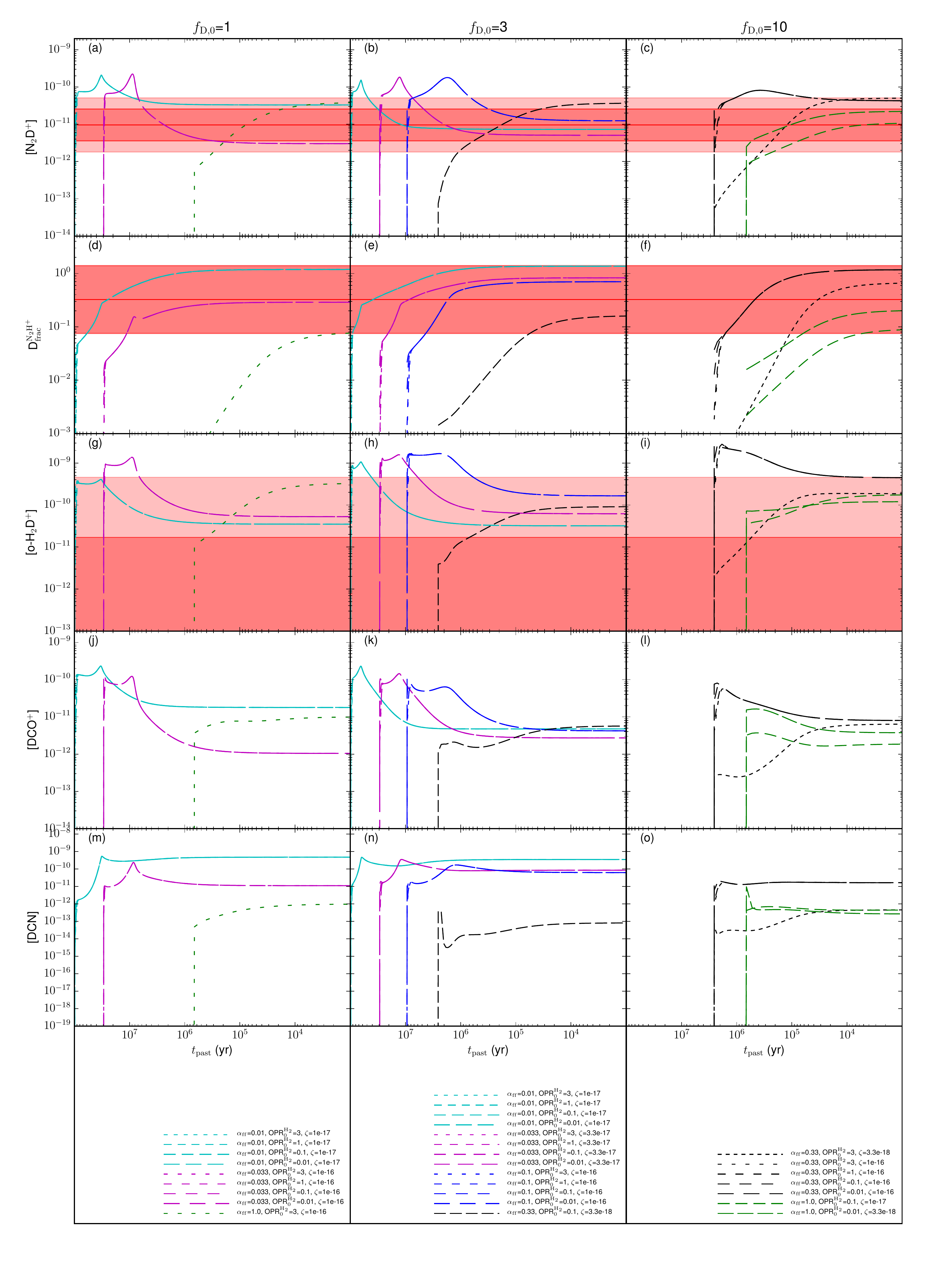}
\caption{
Most promising models for C1-S with $\delta n_{\rm H}^\prime=0.01$, following format of Fig.~\ref{fig:c1ngood10}. 
\label{fig:c1sgood100}}
\end{figure*}

Figures \ref{fig:c1ngood10} and \ref{fig:c1ngood100} show some most
promising models that satisfy the observational constraints for C1-N
with $\delta n_{\rm H}^\prime=0.1$ and 0.01, respectively. Figures
\ref{fig:c1sgood10} and \ref{fig:c1sgood100} show the equivalent
models for C1-S. For C1-S we have included models that fall within a
factor of two of the observational constraints, which allows for
theoretical uncertainties. These figures also show time evolution of
abundances [$\rm DCO^+$] and [DCN], which with future observational
constraints may help to discriminate between the models.

For C1-S with $\delta n_{\rm H}^\prime=0.1$ and focussing on models
with $f_{D,0}\geq3$, we see that the majority of models have
\aff$\leq0.33$. Fast collapse models with \aff=1 require either
relatively low values of \opr (which would imply an already chemical
evolved initial condition) or relatively high values of $\zeta$ or
$f_{D,0}$. Similar conclusions apply to the allowed C1-S models with
$\delta n_{\rm H}^\prime=0.1$, including those starting with no
initial depletion (which may be more reasonable for these lower
initial densities). Improved observational constraints on [\ntdpns],
\dfrac and [\ohtdpns], along with new constraints on [$\rm DCO^+$] and
[DCN], will help to winnow out the allowed models.

For C1-N with $\delta n_{\rm H}^\prime=0.1$ and again focussing on
models with $f_{D,0}\geq3$ we again see that most acceptable models
require relatively small values of \aff. With $\delta n_{\rm
  H}^\prime=0.01$, a small fraction of fast \aff=1 models are allowed,
but these again require low values of \opr, i.e., a chemically
``aged'' initial condition.

\section{Discussion}\label{sec:dc}

\subsection{Implications for Core Dynamics}\label{subsec:dynam}

In order for the cores to be contracting slowly, (\aff $<$ 0.33),
there should be a significant amount of support against gravity. T13
studied the dynamics of C1-N and C1-S. A virial analysis indicated
that they are moderately sub-virial, unless magnetic fields of order
$\sim$ 1~mG are present. Our constraints from the deuteration chemical
clock support this interpretation: a relatively long core history is
needed, otherwise there would not be enough time to reach such high
levels of deuteration (\dfrac $\gtrsim$0.1). Higher resolution
observations that can begin to map \dfrac would provide stronger
constraints, since relative \dfrac values would be more accurately
measured and would constrain the properties of the envelope gas from
which the cores are forming. 

In terms of timescales of the contraction from \nhz to the
  present day, from equation \ref{equ:timedens}, we derive, e.g., for
  C1-S, $t_{\rm past,0}$ = (2.4, 10.0)$\times$10$^5$/\aff yr (for
  $\delta n_{\rm H}^\prime=0.1,0.01$, respectively). In the first
  case, the core age would be $t_{\rm past,0}$ = 7.2$\times$10$^5$ yr
  with \aff = 0.33 and 2.4$\times$10$^6$ yr with \aff = 0.1. These
  timescales are several times longer than some estimates of low-mass
  starless core lifetimes in dense regions from statistics of starless
  cores, protostellar cores and young stellar objects (see review by
  Ward-Thompson et al. 2007), but are comparable with the
  astrochemical estimate of the age of a protostellar core envelope by
  \citet{2014Natur.516..219B}.

%Comparing with studies of low-mass cores by
%\citet{2007prpl.conf...33W} (figure 2), the timescales reported here
%are in the upper limit at similar \nh in their plot, but (much)
%shorter than core ages at lower \nhns. This suggests that core
%density, rather than core mass, is the dominant factor in terms of
%core age.  The $\delta n_{\rm H}^\prime=0.01$ case in our model would
%have longer timescale.  Unfortunately, there is no similar study about
%massive starless cores to compare with.  }

\subsection{\dfrac Measurement in Core}\label{subsec:dfrac}

Given the potential importance of \dfrac as a useful chemical clock,
the measurement of this quantity is worth more attention. Our initial
goal of using multi-transition spectral fitting was to obtain more
accurate measurement of \dfrac than simply using one \ntdp line and
one \nthp line. However, as in our case this can bring more
complexity, especially for \nthp that shows extended emission. The
main sources of uncertainty come from the different spatial scales
that are probed by the observations, with the single dish observations
not resolving the cores (and by varying amounts).

However, \ntdp is less likely to suffer from this problem since it
appears more spatially concentrated: i.e., the localized cores are in
fact defined by their \ntdpns(3-2) emission. Therefore, the \tex derived
from fitting simultaneously the \ntdp lines should be more reliable.
 
Another potential difficulty is that while single-dish measurements
gather the total flux in their beam, the interferometer data is only
sensitive to structures with a specific range of sizes. However, we do
not expect this is a significant problem for at least our {\it CARMA}
and {\it ALMA} data on the C1-N \& S cores. The angular size of both
cores is $\sim$ 7\arcsec.  The angular resolution of {\it ALMA}
observation is 2\arcsec, and the maximum recoverable scale is
9\arcsec. {\it CARMA} observation has 5\arcsec~ synthesized beam, and
the maximum recoverable scale is $\sim$ 50\arcsec.

Determination of [\nthpns] is subject to some ambiguity.  The
locations of C1-N and C1-S are not precisely coincident with \nthp
peaks that are seen in the {\it CARMA} and {\it SMA} maps.
In addition, we see an extended, continuous \nthp structure around
the \ntdp cores. Under such circumstances, it is quite uncertain
what fraction of \nthp line flux is emitted from the \ntdp cores,
and in practice this has been a main contributor to the uncertainty in
the \dfrac measurement.

\citet{2011A&A...529L...7F} have measured \dfrac in high-mass
  starless regions, including C1, to be $\gtrsim$ 0.4, but did not
  resolve the structures. \citet{2012A&A...538A.137M} have measured
  \dfrac as high as $\sim1.0$ in low-mass pre-stellar cores.
  \citet{2013ApJ...765...59F} measured \dfrac in a sample of low-mass
  cores, with mean \dfrac = 0.08 and maximum \dfrac = 0.2. These
  values are comparable to our \dfrac measurements of C1-N and
  C1-S. These suggest that high values of \dfrac ($\gtrsim 0.1$) can
  present in both low-mass and high-mass cores. However, considering
  the shorter free-fall time in high-mass cores (typically they have a
  factor of $\gtrsim$ 10 higher density), the question of how they are
  supported long enough to build up high \dfrac becomes more
  intriguing. As discussed earlier, magnetic fields may play an
  important role here in slowing down their collapse.

\subsection{The Importance of o-H$_2$D$^+$}\label{subsec:oh2dp}

As can be seen from the results, o-H$_2$D$^+$ can place strong
constraints on the modeling. As one of the first products in deuterium
fractionation, o-H$_2$D$^+$ is probably the best observable deuterated
species that gives clues about the progress of deuteration.  For
instance, in the fiducial case (Figures \ref{fig:c1nfancy} and
\ref{fig:c1sfancy}), those models with high enough \dfrac predict
values of [o-H$_2$D$^+$] that are close to the current observational
upper limits. Future, more sensitive observations of o-H$_2$D$^+$
should play a key role in breaking the degeneracies amongst the
currently allowed models.

\section{Conclusions}\label{sec:conc}

We have measured the deuterium fraction \dfrac in two massive
starless/early-stage cores (C1-N and C1-S) first identified by
\citet{2013ApJ...779...96T}. To do this, multiple transitions of
\ntdp and \nthp
lines were observed with {\it ALMA}, {\it CARMA}, {\it SMA}, {\it
  JCMT}, {\it NRO 45m} and {\it IRAM 30m} telescopes. These data
reveal interesting, disturbed kinematics around the cores and also
indicate the presence of significant \nthp emission from the clump
envelope, including a relatively warm component. Still, by considering
a model of emission from the \ntdpns(3-2)-defined cores, excitation
temperatures, \texns, and column densities and abundances of \ntdp
and \nthp in the cores were estimated by simultaneously fitting all
available spectra.

Astrochemical models of collapsing cores have been run with a variety
of initial conditions. The main parameter of our interest is the
collapse rate, \affns. However, results can also depend on the
cosmic-ray ionization rate $\zeta$, initial density \nhz relative to
final density, initial depletion factor \ffdzns, and initial
ortho-to-para H$_2$ ratio \oprzns. Comparison between the observations
and the models suggests the most favorable models have \aff $<$ 0.33
for both C1-N and C1-S, including many models with \aff $\ll1$, so
that there is sufficient time for chemical equilibrium to be
established. The few fast-collapse models that are consistent with the
data require small initial values of \opr, which in itself indicates a
chemcially evolved cloud as the starting condition for core formation.

Our study has shown that measurement of \dfrac and [o-H$_2$D$^+$]
can provide powerful constraints on the dynamics of massive starless/early-stage
cores. However, improved observations, especially of [o-H$_2$D$^+$]
and other deuterated species are needed to disentangle certain
degeneracies amongst the allowed models.

\acknowledgments We thank an anonymous referee for helpful comments.
We thank Jan Wouterloot for helping with the JCMT observation.  SK and
JCT acknowledge an NRAO/SOS grant and NSF grant AST 1411527. PC
acknowledges the financial support of the European Research Council
(ERC; project PALs 320620).  This paper makes use of the following
{\it ALMA} data: ADS/JAO.ALMA\#2011.0.00236.S. {\it ALMA} is a
partnership of ESO (representing its member states), NSF (USA) and
NINS (Japan), together with NRC (Canada), NSC and ASIAA (Taiwan), and
KASI (Republic of Korea), in cooperation with the Republic of
Chile. The Joint {\it ALMA} Observatory is operated by ESO, AUI/NRAO
and NAOJ. Support for {\it CARMA} construction was derived from the
states of California, Illinois, and Maryland, the James S. McDonnell
Foundation, the Gordon and Betty Moore Foundation, the Kenneth T. and
Eileen L. Norris Foundation, the University of Chicago, the Associates
of the California Institute of Technology, and the National Science
Foundation. Ongoing {\it CARMA} development and operations are
supported by the National Science Foundation under a cooperative
agreement (NSF AST 08-38226) and by the {\it CARMA} partner
universities. The James Clerk Maxwell Telescope has historically been
operated by the Joint Astronomy Centre on behalf of the Science and
Technology Facilities Council of the United Kingdom, the National
Research Council of Canada and the Netherlands Organisation for
Scientific Research. We are grateful to the staff members at the
Nobeyama Radio Observatory (NRO) for both operating the 45 m and
helping us with the data reduction; NRO is a branch of the National
Astronomical Observatory, National Institutes of Natural Sciences,
Japan.

{\it Facilities:} \facility{ALMA}, \facility{CARMA}, \facility{SMA}, \facility{JCMT}, \facility{NRO 45m}, \facility{IRAM 30m}.

\end{document}